\documentclass[twocolumn]{aastex631}
\usepackage[mathlines]{lineno}
\usepackage{amsmath}
\usepackage{amssymb}
\usepackage{subfigure}
\usepackage{booktabs}
\usepackage{multirow}
\usepackage{appendix}
\usepackage{url}
\usepackage{color}
\usepackage[T1]{fontenc}

\newcommand{\romaniii}{\uppercase\expandafter{\romannumeral 3}}

\begin{document}

\title{Bright siren without electromagnetic counterpart by LISA-Taiji-TianQin network}

\author[0009-0007-6439-6891]{Yejing Zhan}
\affiliation{School of Astronomy and Space Science, Nanjing University, Nanjing 210093, People’s Republic of China;}

\author{David Izquierdo-Villalba}
\affiliation{Dipartimento di Fisica “G. Occhialini”, Università degli Studi di Milano-Bicocca, Piazza della Scienza 3, 20126 Milano,
 Italy}
\affiliation{INFN, Sezione di Milano-Bicocca, Piazza della Scienza 3, 20126 Milano, Italy}

\author[0000-0001-5174-0760]{Xiao Guo}
\affiliation{Institute for Gravitational Wave Astronomy, Henan Academy of Sciences, Zhengzhou 450046, Henan, China;}
\affiliation{School of Fundamental Physics and Mathematical Sciences, Hangzhou Institute for Advanced Study, University of Chinese Academy of Sciences, No.1 Xiangshan Branch, Hangzhou 310024, China;}

\author[0009-0009-8783-5900]{Qing Yang}
\affiliation{College of Engineering Physics, Shenzhen Technology University, No. 3002 Lantian Road, Shenzhen 518118, China}
\affiliation{Shenzhen Key Laboratory of Ultraintense Laser and Advanced Material Technology, Center for Intense Laser Application Technology, Shenzhen Technology University, No. 3002 Lantian Road, Shenzhen 518118, China}

\author{Daniele Spinoso}
\affiliation{Como Lake Center for Astrophysics, Università degli studi dell'Insubria, via Valleggio 11, 22100, Como, Italy}

\author[0000-0003-4157-7714]{Fa-Yin Wang}
\affiliation{School of Astronomy and Space Science, Nanjing University, Nanjing 210093, People’s Republic of China;}
\affiliation{Key Laboratory of Modern Astronomy and Astrophysics (Nanjing University), Ministry of Education, Nanjing 210093, People's Republic of China;}

\correspondingauthor{Fayin Wang}
\email{fayinwang@nju.edu.cn}


\begin{abstract}
Gravitational waves (GWs) with electromagnetic counterparts (EMc) offer a novel approach to measure the Hubble constant ($H_0$), known as bright sirens, enabling $H_0$ measurements by combining GW-derived distances with EM-derived redshifts. Host galaxy identification is essential for redshift determination but remains challenging due to poor GW sky localization and uncertainties in EMc models. To overcome these limitations, we exploit the ultra-high-precision localization ($\Delta \Omega_s \sim 10^{-4} \, \text{deg}^2$) with a space-based GW detector network (LISA-Taiji-TianQin), which permits unique host identification solely from GW signals. We integrate five massive black hole binary (MBHB) population models and two galaxy number density models to compute the redshift horizon for host galaxy identification and evaluate $H_0$ constraints. We find that (1) The network enhances localization by several orders of magnitude compared to single detectors; (2) The identification horizon reaches $z\sim 1.2$ for specific MBHBs in the most accurate localization case; (3) The population model choice critically impacts the outcomes: the most refined population models yield to independent EMc identification rate of 0.6-1 $\text{yr}^{-1}$ with $H_0$ constraints $< 1\%$ fractional uncertainty, the less refined models lead to the rate $<0.1\text{yr}^{-1}$ and $1-2\%$ uncertainty on $H_0$.  
\end{abstract}




\section{Introduction} \label{sec:intro}


\par The measurement of the Hubble constant $(H_0)$ is a crucial task for validating the standard cosmological model, as a tension has been observed between measurements derived from the local and early universe \citep{verdeTensionsEarlyLate2019,valentinoRealmHubbleTension2021,huHubbleTensionEvidence2023}. The tension may originate from unrecognized measurement error and/or incompleteness of the cosmological model. In particular, early-universe constraints from the cosmic microwave background (CMB) are model-dependent, assuming $\Lambda$CDM cosmology. In contrast, local measurements rely on the redshift-distance relation, in which redshifts depend on the galaxy recessional velocities and the distances are calibrated via the cosmic distance ladder. This method is prone to error accumulation during the calibration of the distance ladder. Beyond systematics, several physical explanations have been proposed, including dynamical dark energy \citep{zhaoDynamicalDarkEnergy2017,jiaUncorrelatedEstimationsH02025,divalentinoCosmoVerseWhitePaper2025,Jia2025}, modified gravity \citep{zumalacarreguiGravityEraEquality2020,odintsovAnalyzingH0Tension2021,petronikolouAlleviatingH0Tension2022}, and local non-isotropy \citep{dingGigaparsecscaleLocalVoid2020,haslbauerKBCVoidHubble2020}. Thus, an independent measurement on $H_0$ is essential to adjudicate the tension.

\par Gravitational waves (GWs) with electromagnetic counterpart (EMc) offer a novel approach to measure the $H_0$, independently, known as the bright sirens method, with GW providing a luminosity distance by analyzing the waveform and EMc providing the redshift by its identified host galaxy. \citep{schutzDeterminingHubbleconstant1986,holzUsingGravitationalWaveStandard2005,abbottGravitationalwaveStandardSiren2017}. The first bright siren detected in 2017, composed of a GW event (GW170817) and a gamma-ray burst (GRB 170817A), constrains the   
$H_0$ to be $70^{+12.0}_{-8.0} \text{km s}^{-1}\text{Mpc}^{-1}$ \citep{ligoscientificcollaborationandvirgocollaborationGW170817ObservationGravitational2017,abbottGravitationalwaveStandardSiren2017}. However, no subsequent bright sirens have been detected by ground-based GW observatories, largely due to poor sky localization ($\Delta \Omega_s\sim 100-1000 \text{ deg}^2$)\citep{groverComparisonGravitationalWave2014,chassande-mottinGravitationalWaveObservations2019} and low rate of joint detection of GWs and EMc \citep{Zhang2018,chruslinskaDoubleNeutronStars2018}. Poor sky localization complicates EMc identification by generating numerous host galaxy candidates. The space-based detectors improve the localization to $\sim 10\text{ deg}^2$ for the sake of much larger detection arm lengths
\citep{babakScienceSpacebasedInterferometer2017,ruanTaijiProgramGravitationalwave2020,mangiagliObservingInspiralCoalescing2020,piroChasingSupermassiveBlack2023,colpiLISADefinitionStudy2024,liGravitationalWaveAstronomy2025}, which provides a more accurate localization but still leads to millions of host galaxy candidates \citep{lopsGalaxyFieldsLISA2023}. In addition, the EMc models remain uncertain, though previous studies have researched deeply in this realm \citep{holzUsingGravitationalWaveStandard2005,kocsisPremergerLocalizationGravitational2008,bellovaryMIGRATIONTRAPSDISKS2016,cowperthwaiteElectromagneticCounterpartBinary2017,bogdanovicElectromagneticCounterpartsMassive2022,jinTaijiTianQinLISANetworkPrecisely2023,yeObservingWhiteDwarf2023,chenElectromagneticCounterpartsPowered2024,zhanPreciseHubbleconstant2025}, which may result in overestimating or underestimating the event rate of the bright sirens.

\par The redshift can also be extracted by statistically matching their sky localization region with mock or real galaxy catalogs, without identifying the host galaxy of a single source. This method is called the `dark siren' method, which was first brought up by \cite{schutzDeterminingHubbleconstant1986}. This method is well tested with real data from the ground-based GW detectors, though the constraint on $H_0$ by this method is not very competitive with a fractional error of $\sim 20-40\%$ on $H_0$ \citep{grayJointCosmologicalGravitationalwave2023,bomDarkStandardSiren2024}. Some simulations predict that the measurement on $H_0$ by the dark siren method could be enhanced to a few-percentage level in the future \citep{macleodPrecisionHubbleconstant2008,chenTwoCentHubble2018,laghiGravitationalwaveCosmologyExtreme2021,jinTaijiTianQinLISANetworkPrecisely2023,zhuDarkSideUsing2023,songSynergyCSSTGalaxy2024}. However, $H_0$ can be biased relative to the true value, even with optimistic assumptions. This bias stems from incorrect galaxy redshift distributions or preferential weighting of incorrect host galaxies \citep{trottChallengesStatisticalGravitationalwave2023,hanselmanGravitationalwaveDarkSiren2025}.  

\par There is a growing interest in figuring out redshift information by the GW signal alone. Novel methods are developed to estimate the redshift via the GW signal, including but not limited to inferring the redshift from the equation of state of the neutron star binaries, estimating the redshift of the host galaxies via the mass-luminosity relations of the galaxies, e.t.c. \citep[see the references for detail;][]{taylorCosmologyUsingAdvanced2012,namikawaAnisotropiesGravitationalWaveStandard2016, delpozzoCosmologicalInferenceUsing2017,zhuConstrainingCosmologicalParameters2022,ezquiagaSpectralSirensCosmology2022}. With these methods, a promising one involves identifying the unique host galaxy by high-precision GW localization. 

\par For a space-based GW detector, the localization of the GW signal provides the distance and sky angle location of a source with uncertainty, then forming an uncertainty volume. If the volume is small enough to ensure fewer than one host galaxy candidate, one could identify the host galaxy without specific EMc.  Given a typical sky location uncertainty of $\sim 10 \text{ deg}^2$, the typical distance uncertainty of $\sim 10\%$ \citep{gongConceptsStatusChinese2021,colpiLISADefinitionStudy2024,liGravitationalWaveAstronomy2025} and a typical galaxy number density of $0.02 \text{ Mpc}^{-3}$ \citep{blantonGalaxyLuminosityFunction2003}, one can identify the host galaxies in the local universe ($<100 \text{Mpc}$). \cite{kyutokuGravitationalwaveCosmographyLISA2017} has discussed the prospects for the identification of the host galaxies in the local universe via the stellar-mass black hole binary mergers with single detectors. Typically, the sky location measurement for a single detector lies in response to the detector using the arrival time differences of GW between constituent satellites of the detector. In contrast, multiple detectors compose a network, significantly improving localization \citep{ruanLISATaijiNetwork2020,zhangSkyLocalizationSpacebased2021}. The network utilizes the arrival time differences between different detectors, which are several orders of magnitude larger than those between satellites, leading to high-precision localization. Thus, the detector network provides strong potential for identifying the host galaxy without EMc.   

\par Three space-based GW detectors are expected to launch in the 2030s, including Laser Interferometer Space Antenna \citep[LISA;][]{amaro-seoaneLaserInterferometerSpace2017}, Taiji \citep{huTaijiProgramSpace2017}, Tianqin \citep{luoTianQinSpaceborneGravitational2016}. If these space-based detectors have a significant time overlap in the 2030s, the detector network could lead to significantly enhancing the localization of the GW sources \citep{ruanLISATaijiNetwork2020,wangAlternativeLISATAIJINetworks2021,shumanMassiveBlackHole2022}, hence increasing the possibility of identifying host galaxies without EMc. For example, \cite{ruanLISATaijiNetworkPrecision2021} pointed out that a massive black hole binary (MBHB) could identify its host galaxy within a redshift range of $0.5-0.9$ by the LISA-Taiji network.  

In this work, we combine the capabilities of the LISA-Taiji-TianQin network to assess the detectability of MBHBs whose host galaxies can be identified solely with the GW signal. Since the detectability of these systems is expected to depend on both the galaxy number density and the MBHB population, we consider five different MBHB population models and two distinct galaxy number density scenarios.
We calculate the maximum redshifts for host galaxy identification for different MBHB with specific parameters. In addition, we calculate detection rates of the MBHB with host identification for population models and GND models. The detections are utilized as bright sirens to constrain $H_0$ under the standard cosmological model for each population and galaxy number density model. 
\par We summarize the method of host galaxy identification solely using GW in Sect. \ref{sec:method}, including the MBHB population models described in Sect. \ref{sec:population}, the galaxy number density models described in Sect. \ref{sec:GND} and the GW detection described in Sect. \ref{sec:GWdetection}. The result of the detection rates and the constraint on $H_0$ by this method are presented in Sect. \ref{sec:result}. Specifically, we discuss the enhancement of employing the detector network compared to single detectors in Sect. \ref{sec:improvement}. Finally, Sect. \ref{sec:conclusion} summarizes the main findings.

\section{Method}\label{sec:method}

\par The identification of unique host galaxies for MBHBs through GW depends critically on three decisive factors. First, the galaxy number density plays a fundamental role, as a compact galaxy density results in a large number of host galaxy candidates. Second, the localization capabilities of GW detectors directly determine the size of the sky region within which host galaxy candidates must be searched. Third, the adopted MBHB population model significantly influences the expected redshift and mass distributions of MBHBs, leading to impacts on localization.
\par In this section, we describe our methodology for quantifying these factors and present the framework used to evaluate the prospects for host galaxy identification. We also specify the cosmological model adopted in our analysis.

\subsection{MBHB populations}\label{sec:population}
\par Modeling the MBHB population requires reconstructing the hierarchical merger history of dark matter halos and the initial seeding of black holes. The merger history could be derived from the cosmological \textit{N}-body simulation, such as \texttt{Millennium} and \texttt{Millennium-\uppercase\expandafter{\romannumeral 2}}  \citep{springelSimulationsFormationEvolution2005,boylan-kolchinResolvingCosmicStructure2009}. Alternatively, such histories could be described by analytical methods, such as the extended Press-Schechter (EPS) formalism, which successfully reproduces the result of the \textit{N}-body simulation \citep{pressFormationGalaxiesClusters1974,coleHierarchicalGalaxyFormation2000,volonteriAssemblyMergingHistory2003}. The prevailing seeding models are separated into two kinds, i.e., the ``light-seeding" model and the ``heavy-seeding" model. The ``light-seeding" model assumes that the initial black holes with a mass around $300 M_\odot$ are seeded in the remnants of the population \romaniii~ stars at a high redshift of 15-20 \citep{latifFormationSupermassiveBlack2016,inayoshiAssemblyFirstMassive2020}, while the ``heavy-seeding" model assumes that the seeds come from the direct collapse of the gas, forming massive black holes with masses of $\sim 10^5 M_\odot$ at similar redshifts \citep{volonteriBlackHolesEarly2012,mayerRouteMassiveBlack2018}. 

\begin{figure*}[htbp]
    \centering
    \includegraphics[width=1.\linewidth]{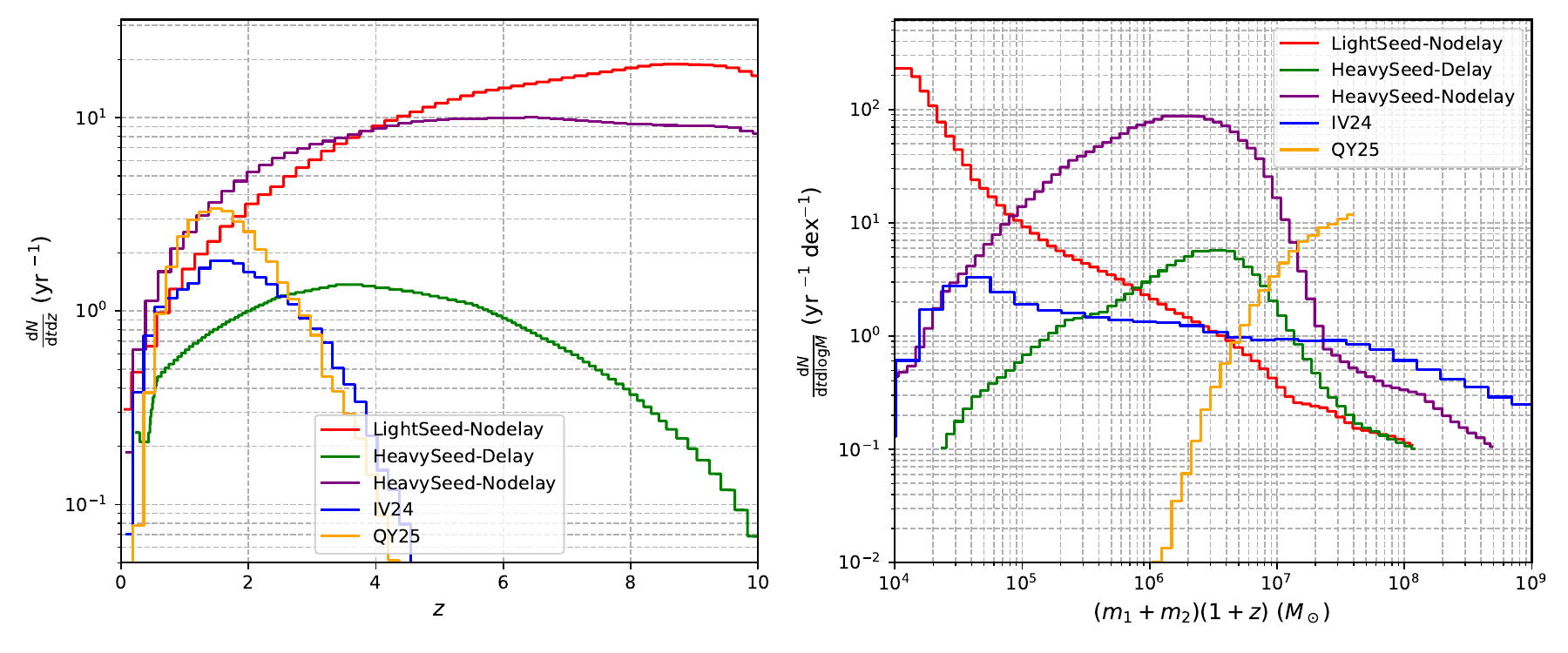}
    \caption{The merger rate of all population models  across varying redshift (left) and binary mass (right). The MBHB population at high redshift ($z>4$) is predominated in the LightSeed-Nodelay and HeavySeed-Nodelay models, whereas the IV24 and QY25 models predict a significant MBHB population at low redshifts ($z < 4$). In the HeavySeed-Delay model, the population is largely uniformly distributed in a redshift range of 1 to 6. Regarding binary mass, the LightSeed-Nodelay model predicts a prevalence of low-mass MBHBs with total masses below $10^5 M_\odot$, while the HeavySeed-Delay and HeavySeed-Nodelay models show a dominance of MBHBs with masses in the range $10^6-10^7M_\odot$. The mass distribution of IV24 is largely uniform in the range of $10^4-10^8 M_\odot$. The QY25 model favors high-mass MBHBs with mass over $10^6M_\odot$. 
    }  
    \label{fig:population}
\end{figure*}

\par In this work, we focus on five population models. Three models construct the merger tree by the EPS formalism with different seeding models:
\begin{enumerate}
    \item LightSeed-Nodelay model: ``Light-seeding" from  population \romaniii~stars with a mass sampled from a log-normal distribution centered at $300 M_\odot$. The black hole is seeded in the halos at $3.5\sigma$ peaks of the primordial density field in the redshift range $15<z<20$. This model omits the time delay between the galaxy mergers and the MBHB mergers.
    \item HeavySeed-Delay model: ``Heavy-seeding" with a mass around $10^5 M_\odot$ from the collapse of protogalactic disks. The seeding process is determined by the critical Toomre parameter for disk instability \citep{volonteriEvolutionMassiveBlack2008}. This model incorporates a time delay between the galaxy mergers and the MBHB mergers, accounting for the environmental effects including tidal stripping, tidal evaporation and dynamical friction \citep{barausseEvolutionMassiveBlack2012}. 
    \item HeavySeed-Nodelay model: identical to the HeavySeed-Delay model, but without the time delay between the galaxy mergers and MBHB mergers.
\end{enumerate}

These models are well-simulated in \cite{kleinScienceSpacebasedInterferometer2016}, who simulate the catalogs with halo masses ranging from $10^{10}M_\odot$ to $10^{16} M_\odot$. We sample these three models to catalogs according to the event rates by a toolbox, \texttt{GWtoolbox} \citep{yiGravitationalWaveUniverse2022}, which contains flexible tools to simulate observations of GW with different detectors.

\par In addition, we also consider two more refined models extracting the merger tree from the dark matter \textit{N}-body simulation. Generally, these simulation computationally models the gravitational dynamics of billions of dark matter particles to reconstruct the hierarchical formation of cosmic structures, such as halos that host galaxies and enable the growth of massive black holes via mergers and accretion processes. The two models are described as follows:
\begin{enumerate}
    \setcounter{enumi}{3}
        \item IV24 model: uses a variant of the L-Galaxies semi-analytical galaxy formation model, which was calibrated on top of the merger trees of both the \texttt{Millennium} and \texttt{Millennium-\uppercase\expandafter{\romannumeral 2}} simulation, as shown in \cite{izquierdo-villalbaGalacticNucleiHalo2020,spinosoMultiflavourSMBHSeeding2022,izquierdo-villalbaConnectingLowredshiftLISA2024}. This simulation covers a volume of $(100 \mathrm{Mpc/h})^3$ and can resolve structures ranging from $10^8 M_\odot/\mathrm h$ to approximately $3\times 10^{14} M_\odot/\mathrm h$ \citep{boylan-kolchinResolvingCosmicStructure2009}. Regarding the formation of the first MBHs, the model incorporates a multi-flavor seeding approach (light $\sim 100M_\odot$, medium $\sim 10^3 M_\odot$ and heavy $\sim 10^5 M_\odot$) based on environmental properties such as the Lyman-Wernel flux and metal pollution \citep{spinosoMultiflavourSMBHSeeding2022}. Additionally, it features a sophisticated model to account for the dynamical evolution of MBHBs following galaxy-galaxy mergers \citep{izquierdo-villalbaMassiveBlackHole2022}.
    \item QY25 model: it utilizes the L-Galaxies semi-analytical framework \citep{guoDwarfSpheroidalsCD2011,guoGalaxyFormationWMAP12013} built on the \texttt{Millennium} simulation, which covers a volume of $(500 \mathrm{Mpc/h})^3$ and can resolve dark matter structures ranging from $10^{10} \mathrm{M_{\odot}}$ to approximately $10^{16} \mathrm{M_{\odot}}$. MBHs grow in the center of dark matter halos under both `quasar' and `radio' modes. For MBHBs, the QY25 model adopts similar MBHB construction methods to \cite{yangAnisotropyNanohertzGravitationalwave2025} and generates an MBHB catalog by selecting light-cone events between adjacent redshift snapshots, considering the source-to-Earth propagation time while omitting the time delay between MBHB mergers and their host galaxies.
\end{enumerate}

The models predict event rates of 166, 8, 120 mergers per year for LightSeed-Nodelay, HeavySeed-Delay and HeavySeed-Nodelay models, respectively, and 6, 6 mergers per year for IV24, QY25 models, respectively. Notably, the rates presented in the text represent the merger rates in the whole universe but not the detection rates. For instance, a large portion of the MBHB merger could not be detected by the GW detectors due to an incompatible frequency range, which means the detection rate may be significantly lower than the event rate. 

\par The merger rates for the five population models across redshifts and binary masses are illustrated in Fig. \ref{fig:population}. The merger rates indicate that the MBHB population is predominated at high redshift $z \gtrsim 4$ for the LightSeed-Nodelay and HeavySeed-Nodelay models. The population is largely uniformly distributed over the range of 1 to 6 in the HeavySeed-Delay model. In contrast, the IV24 and QY25 predict a significant MBHB population at low redshifts ($z < 4$), with a peak at $z\sim 1.6$. Regarding binary mass, the LightSeed-Nodelay predicts a prevalence of low-mass MBHBs with total masses below $10^5 M_\odot$, while the HeavySeed-Delay and HeavySeed-Nodelay models show a dominance of MBHBs with masses in the range $10^6-10^7M_\odot$. The mass distribution of IV24 is largely uniform in the range of $10^4-10^8 M_\odot$. The QY25 model favors high-mass MBHBs with mass over $10^6M_\odot$. 


\subsection{The galaxy number density}\label{sec:GND}

\par The galaxy number density ($n_{\text{gal}}$, GND) can be simply assumed as uniform, as employed in some works \citep{kyutokuGravitationalwaveCosmographyLISA2017,ruanLISATaijiNetworkPrecision2021,jinTaijiTianQinLISANetworkPrecisely2023}. However, the GND changes with the redshift, with a decreasing trend towards higher redshifts \citep{conseliceEvolutionGalaxyNumber2016,agazieNANOGrav15yearData2023}. In this work, we employ two GND models: 
\begin{itemize}
    \item uniform GND: a constant value of $n_\text{gal}=0.02 \text{ Mpc}^{-3}$, consistent with typical estimations \citep{blantonGalaxyLuminosityFunction2003}.
    \item $z$-dependent GND: A model where GND evolves with redshift, as described below. 
\end{itemize}
The $z$-dependent GND can be described by the stellar mass function $\Phi(M,z)$, representing the number density in the intervals of stellar mass $[M,M+\Delta M]$ at the redshift interval $[z,z+\Delta z]$. The stellar mass function is empirically expressed by the Schether function \citep{schechterAnalyticExpressionLuminosity1976}, parameterized by the characteristic mass $M_*$, power index $\alpha$, and the normalization factor $\phi_*$. We adopt the expanded two-component function \citep{pengMASSENVIRONMENTDRIVERS2010}, characterized by five parameters $M_*,\phi_1,\alpha_1,\phi_2,\alpha_2$
\begin{equation}
\begin{aligned}
        n_\text{gal}(M,z)\mathrm d &M=\Phi(M,z) \mathrm d M\\
        =&\left[\phi_1^*\left(\frac{M}{M^*}\right)^{\alpha_1} e^{-M/M^*}+\right.\\
        &\left. \phi_2^*\left(\frac{M}{M^*}\right)^{\alpha_2} e^{-M/M^*}\right]_z\mathrm d \left(\frac{M}{M^*}\right)_z
\end{aligned}
\label{eq:schether}
\end{equation}
where the subscription $z$ represents the parameters are in the value at the redshift $z$. We follow the fitting result of the parameters from \cite{weaverCOSMOS2020GalaxyStellar2023}. We define a notation for the GND in a stellar mass range:
\begin{equation}
    n_{a,b}(z)=\int^b_a n_\text{gal}(M,z) \mathrm d M
    \label{eq:galaxy-number-density}
\end{equation}
For example, $n_{8,9}(z)$ represents the number density of the galaxies in the stellar mass range between $10^8 M_\odot$ and $10^{9} M_\odot$. 

\begin{figure}[htbp]
    \centering
    \includegraphics[width=1.\linewidth]{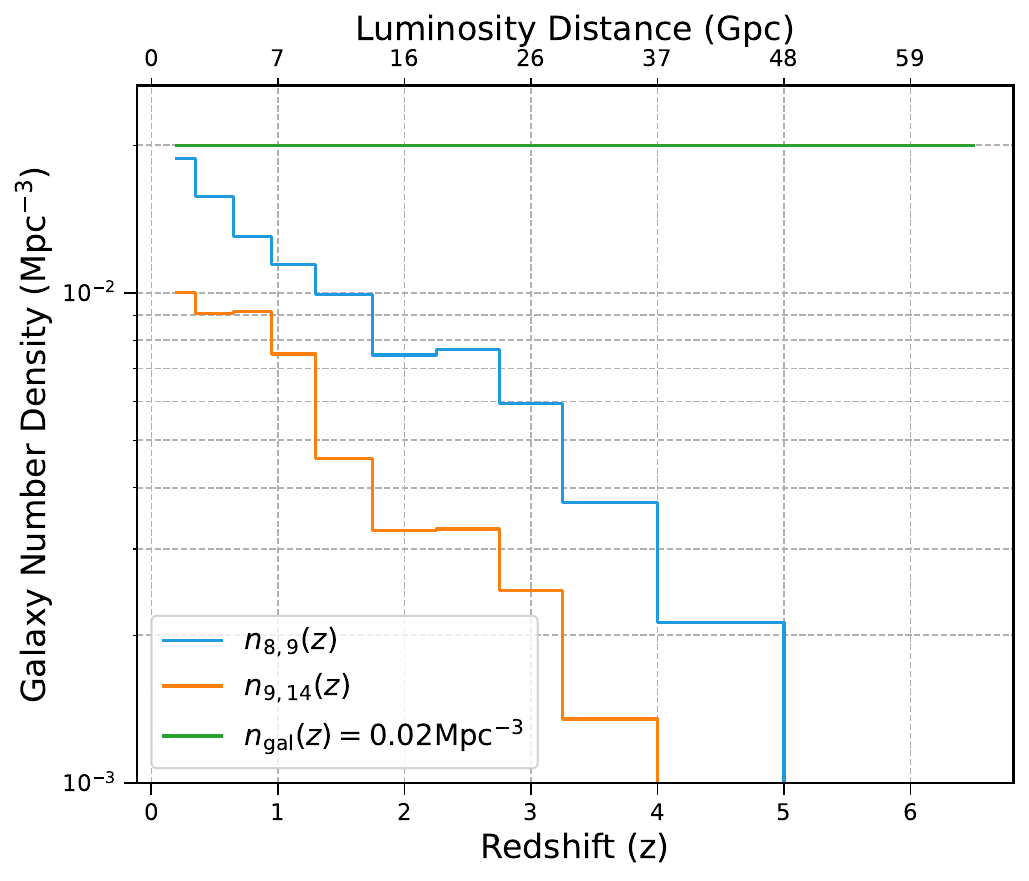}
    \caption{The GND in different models as the redshift changes. The blue line represents the filter in the mass range of $10^8-10^{9} M_\odot$, defined by Eq. (\ref{eq:galaxy-number-density}), while the orange line represents the filter in the mass range of $10^9-10^{14} M_\odot$.}
    \label{fig:GND}
\end{figure}

\par The observation shows that the mass of the central black hole scales with the total stellar mass of the galaxy with a factor of $\sim 0.1\%$ in the dwarf galaxy (stellar mass lower than $10^9M_\odot$) \citep{reinesRELATIONSCENTRALBLACK2015,schutteBlackHoleBulge2019}, and the factor would be increased to $\sim 0.5\%$ for the larger galaxies \citep{magorrianDemographyMassiveDark1998,kormendyCoevolutionNotSupermassive2013,davisBlackHoleMass2019}. We leverage this relation to filter galaxy candidates in two components, adopting:
\begin{enumerate}
    \item Stellar mass range  $10^8-10^{9} M_\odot$ for MBHBs with masses $10^5<m_1+m_2<10^6M_\odot$. The GND in this range is denoted by $n_{8,9}(z)$.
    \item Stellar mass range $10^9-10^{14} M_\odot$ for MBHBs with masses $m_1+m_2>10^6 M_\odot$, in which the density is denoted by $n_{9,14}(z)$
\end{enumerate}
where $m_1,m_2$ represent the black hole masses in the MBHB.

\par The GND for both components is shown in Fig. \ref{fig:GND}. Furthermore, the high GND of low-mass galaxies reduces the likelihood of detection for binaries hosted in such systems. Thus, we choose a mass cutoff for the MBHB population in this work, requiring $m_1>10^5 M_\odot$ or $m_2>10^5 M_\odot$ to exclude low-mass dwarf galaxies.

\subsubsection{Cosmological models}
\par To trace the expansion history in high redshift where the MBHB populations reside, we adopt the standard $\Lambda$CDM model. Assuming a flat universe ($\Omega_m+\Omega_\Lambda=1$), the distance-redshift relation is   
\begin{equation}
    d_L(z)=\frac{c(1+z)}{H_0}\int \frac{\mathrm d z}{\sqrt{\Omega_\Lambda+\Omega_{m}(1+z)^3}}
\end{equation}
where $c$ is the speed of light, $\Omega_\Lambda$ is the dark energy density parameter and $\Omega_m$ is the matter density parameter. Cosmological parameters are constrained by \textit{Planck}+2015 results \citep{adePlanck2015Results2016}, where $\Omega_m=0.315$ and $H_0=100h \text{ km s}^{-1} \text{Mpc}^{-1}=67.31 \text{ km s}^{-1} \text{Mpc}^{-1}$ and $h$ is the reduced Hubble constant.

\subsection{GW detectors}
Three space-based GW detectors are scheduled to launch in 2030s. These missions are designed to be sensitive to millihertz-frequency gravitational waves, covering a range of approximately $0.01-1\text{ Hz}$, making them ideally suited for detecting mergers of MBHB mergers. The brief introduction of these three missions is provided below:
\begin{itemize}
    \item LISA: A Europe-led mission. LISA will consist of three satellites arranged in an equilateral triangle with arm lengths of $2.5\times 10^9 \text{ m}$ \citep{amaro-seoaneLaserInterferometerSpace2017}. The mission is planned for launch in 2035 and is designed to operate for 4.5 years, with consumables allocated to support a potential extension to ten years \citep{colpiLISADefinitionStudy2024}.
    \item Taiji: A Chinese-led mission. Taiji adopts a configuration similar to LISA but features a longer arm length of $3\times 10^9 \text{ m}$. The mission is expected to launch in the early 2030s with a nominal operational lifetime of five years \citep{huTaijiProgramSpace2017}.
    \item TianQin: A Chinese-led mission. The configuration of TianQin is similar with LISA, but with a shorter arm length of $\sim 1.7\times 10^8\text{ m}$. TianQin is designed for a five-year lifetime, launched in the mid-2030s \citep{luoTianQinSpaceborneGravitational2016}.
\end{itemize}
Based on the projected launch schedules and nominal lifetimes, we assume a four-year period during which all three detectors will operate simultaneously. This overlap will enable the formation of a space-based GW detector network, significantly enhancing sky localization accuracy.

\subsection{The GW detection}\label{sec:GWdetection}
\subsubsection{Waveform modeling} \label{sec:waveform}

In this study, we adopt a circular orbit approximation for MBHBs. Specifically, we adopt the sophisticated phenomenological model \texttt{IMRphenomenD} model, which is designed for non-precessing systems on circular orbit \citep{khanFrequencydomainGravitationalWaves2016,husaFrequencydomainGravitationalWaves2016}. This model provides the waveform in (2,2) harmonic mode, which dominates in such a system.

\par The waveform of \texttt{IMRphenomenD} model depends on seven parameters $\Xi_0\in (m_1,m_2,\chi_1,\chi_2,D_L,\iota,\psi)$, including the binary masses $(m_1,m_2)$, the dimensionless spin $(\chi_1,\chi_2)$, the luminosity distance $(D_L)$, the inclination angle $(\iota)$ and the polarization angle $(\psi)$. The waveform is separated into three frequency regions, namely, the inspiral (Ins), the intermediate (Int), and the merger/ringdown (MR) region:

\begin{equation}
    \tilde h(\Xi_0;f)=
    \left\{\begin{aligned}
        &\mathcal{A}_{\text{Ins}}(\Xi_1;f)e^{i\Psi_{\text{Ins}}(\Xi_2;f)},\quad f<f_1\\
        &\mathcal{A}_{\text{Int}}(\Xi_1;f)e^{i\Psi_{\text{Int}}(\Xi_2;f)}, \quad f_1<f<f_2\\
        &\mathcal{A}_{\text{MR}}(\Xi_1;f)e^{i\Psi_{\text{MR}}(\Xi_2;f)} \quad f>f_2
    \end{aligned}\right.
\end{equation}
where $\Xi_1\in (m_1,m_2,D_L,\iota)$, $\Xi_2\in (m_1,m_2,\chi_1, \chi_2,\psi)$ and $f_1, f_2$ are the characteristic frequency to separate the three regions \citep{khanFrequencydomainGravitationalWaves2016}.

\par The detection of GW relies critically on the response of the detectors. Due to the detector motion with respect to the GW source, corrections to the detector response are necessary. In some cases, the response could be simplified using long-wavelength approximation, when the GW wavelength ($\lambda$) is larger than the detector size $L$, namely, $\lambda/(2\pi)>> L$. This approximation treats the GW field as uniform across the detector, assuming a constant response over the observation time. For the space-based detectors with the detector size $L\sim  10^9\text{ m}$, the GW wavelength for the approximation should be $ \sim  10^{10}\text{ m}$, corresponding to a frequency $\lesssim 20 \text{ mHz}$. 

\par For MBHB systems, the validity of the long-wavelength approximation depends on the evolutionary phase being observed. During the inspiral phase, where the MBHB acts as a continuous GW source, the characteristic GW frequency is much lower than 20 mHz, making the long-wavelength approximation applicable. However, during the merger phase, which we focus on in this work and where the MBHB produces a GW burst, the long-wavelength approximation is not valid. The signal-to-noise ratio (SNR) accumulation for GW signals depends on the binary black hole mass. For the binary mass lower than $10^6 M_\odot$, SNR is dominated by the inspiral phase, accumulating primarily in the final months before merger. In contrast, the SNRs are dominated by the merger and ringdown phase for the system with mass above $10^6M_\odot$ \citep{flanaganMeasuringGravitationalWaves1998,katzEvaluatingBlackHole2019}. In both scenarios, the overall SNR of the MBHBs is largely determined by the final stages of MBHB evolution. These stages are characterized by a frequency around $1-10 \text{ mHz}$, where the long-wavelength approximation is invalid. Therefore, a time-dependent response function, accounting for detector motion, is essential. 
\par Theoretically, GW detectors are designed to be equal-arm to ensure that laser noise experiences the same delay in each arm, allowing it to be canceled using the phase difference. However, due to the motion of the detectors, the arms become unequal in length over time. To address this, the time-delay interferometry (TDI) technique is employed, which constructs virtual equal-length optical paths with different combinations of signal components from the detector (see \cite{vallisneriSyntheticLISASimulating2005,tintoTimedelayInterferometry2020} for details). 

In this work, we adopt the first generation of TDI Michelson-like $(X,Y,Z)$ channels, which depend on the GW strain $\tilde{h}$, the sky angle location $(\theta_s, \phi_s)$, and the coalescence time $t_c$. The $(X,Y,Z)$ channels could be combined linearly to form the optimized orthogonal $(A,E,T)$ channels:
\begin{equation}
    \begin{aligned}
 & A=\frac{1}{\sqrt{2}}(Z-X), \\
 & E=\frac{1}{\sqrt{6}}(X-2Y+Z), \\
 & T=\frac{1}{\sqrt{3}}(X+Y+Z).
\end{aligned}
\label{eq:tdi}
\end{equation}
where the channels $A, E$ represent the two polarizations, and the channel $T$ represents the noise. The time-dependent detector response and the TDI combinations are computed using the \texttt{GWspace} toolbox \citep{liGWSpaceMultimissionScience2025}.

\subsubsection{Parameter estimation: Fisher matrix} \label{sec:parameter_estimation}

\par Identifying the host galaxy of the GW source is one of key tasks in GW cosmology. The standard approach involves searching for EMc within the sky localization region derived from GW observations. The sky location could be measured by the response of the detector, using the GW arrival time difference between detector arms of a single detector, where longer arms provide higher timing precision. In this case, space-based detectors, with arm lengths of $\sim 10^9$ m, offer significant advantages over ground-based detectors ($\sim 10^3$ m) for sky localization. Alternatively, employing a network of detectors significantly improves localization by utilizing time delays \textit{between} detectors. For example, the sky location of the first detected neutron star binary merger GW170817 is improved from $190 \text{ deg}^2$ using only the ground-based GW detector LIGO (Hanford-Livingston observatory) to $28\text{ deg}^2$ analyzed by the LIGO-Virgo network \citep{ligoscientificcollaborationandvirgocollaborationGW170817ObservationGravitational2017}. Naturally, the space-based detector network will provide exceptionally precise sky localization due to the arm length, benefiting the identification of the host galaxy. If the localization is sufficiently precise, the host galaxy could be uniquely identified without EMc.

\par The GW localization relies on parameter estimation of the GW signal. The parameter estimation of waveform parameters bases on the match-filter method, which is described by the inner product of the signal. The inner product is defined by
\begin{equation}
    (h|g)_k=2\Re \left[\int \frac{h(f)g(f)^*+h(f)^*g(f)}{S_{n;k}(f)}\mathrm d f\right]
\end{equation}
where the subscript $k$ represents the result from the $k$-th detector, i.e., $S_{n,k}(f)$ is the one-sided noise spectrum of the $k$-th detector. Both polarization components are used. The SNR for the detector network is defined by
\begin{equation}
    \rho=\sqrt{\sum_k \rho^2_k}=\sqrt{\sum_k (\tilde h|\tilde h)_k}
\end{equation}
where $\rho_k$ is the SNR for a single detector.

\par The Fisher information matrix (FIM) is commonly featured in estimating the uncertainty on the parameters instead of completely Bayesian sampling methods. The uncertainty calculated by the FIM is consistent with the Bayesian method if the signal is observed at a high enough SNR \citep{vallisneriUseAbuseFisher2008}. Notably, the GW in our case is detected at a very high SNR (>100) for the localization of the host galaxy, which ensures that the FIM is valid. 

\par As it is described in Sect. \ref{sec:waveform}, the waveform model depends on ten parameters, i.e. $\Xi\in (m_1,m_2,\chi_1, \chi_2,D_L,\iota, \psi, \theta_s, \phi_s, t_c)$. 
For the $k$-th detector, the Fisher information matrix is 
\begin{equation}
    \Gamma_{ij;k}=\left( \frac{\partial \tilde h}{\partial \Xi_i} \middle | \frac{\partial \tilde h_k}{\partial \Xi_j}\right)_k
\end{equation}
where subscripts $i,j$ represent \textit{i}-th and \textit{j}-th parameters. The FIM for a detector network is the summation of the FIM of the single one, namely 
\begin{equation}
    \Gamma_{ij}=\sum _k \Gamma_{ij;k}
\end{equation} 

\par The covariance of the parameters can be represented by the FIM, i.e., $\text{Cov}_{ij}(\Xi)=1/\Gamma_{ij}(\Xi)$. So the measurement error of the parameters should be $\Delta { \Xi_i}=\sqrt{\text{Cov}_{ii}(\Xi)}=1/\sqrt{\Gamma_{ii}(\Xi)}$. In addition, the distance measurement is also affected by the weak lensing effect due to the gravitational lensing from structures along the line of sight, which leads to uncertainties on the distance. We adopt the fitting formula for the uncertainty \citep{hirataReducingWeakLensing2010}:
\begin{equation}
\Delta_{\text {lens}}d_L=d_{L} \times 0.066\left(\frac{1-(1+z)^{-0.25}}{0.25}\right)^{1.8}    \label{eq:weak-lensing}
\end{equation}

\par An additional source of distance uncertainty arises from the peculiar velocity of the host galaxy, which is dominated at the redshift $\lesssim 0.1$ \citep{heAccurateMethodDetermine2019}. Under the low-redshift approximation, this contribution can be derived from Hubble's law as
\begin{equation}
    \Delta_\text{pv} d_L\simeq (1+z) \frac{\langle{v}\rangle}{H_0}
    \label{eq:dL_pv}
\end{equation}
where $\langle{v}\rangle$ is the mean peculiar velocity at the low redshift.  This velocity is set as a typical value of $\langle{v}\rangle =300 \text{km/s}$ \citep{adePlanckIntermediateResults2014,carrickCosmologicalParametersComparison2015,grazianiPeculiarVelocityField2019}.

\par Therefore, the total distance uncertainty is the combination of both uncertainties:
\begin{equation}
\begin{aligned}
    \Delta_\text{tot} d_L& =\sqrt{(\Delta_\text{lens}d_L)^2+(\Delta_\text{pv}d_L)^2+(\Delta d_L)^2}\\
    &=\sqrt{(\Delta_\text{lens}d_L)^2+(\Delta_\text{pv}d_L)^2+\frac{1}{\Gamma_{d_Ld_L}}}
\end{aligned}
\end{equation}
where $\Delta d_L = 1 / \sqrt{\Gamma_{d_L d_L}}$ is the instrumental uncertainty derived from the FIM.

Specifically, the sky angle localization uncertainty is defined by
\begin{equation}
    \Delta\Omega_s=2\pi|\sin\theta_s|\sqrt{\langle\Delta\theta_s^2\rangle\langle\Delta\phi_s^2\rangle-\left\langle\Delta\theta_s\Delta\phi_s\right\rangle^2}
    \label{eq:skylocation}
\end{equation}
where $\Delta\theta_s,\Delta\phi_s$ are the uncertainties on the sky location, respectively, and $\left\langle\Delta\theta_s\Delta\phi_s\right\rangle=1/\sqrt{\Gamma_{\theta_s\phi_s}(\Xi)}$ is the covariance.

The volume uncertainty can be calculated by
\begin{equation}
    \Delta V=d_L^2\Delta_\text{tot} d_L\Delta \Omega_s 
    \label{eq:volume-uncertainty}
\end{equation}
The host galaxy candidates of the MBHBs should fall within the volume uncertainty. Specifically, a host galaxy is uniquely identified when the number of candidates is one or fewer. In this work, MBHB detection is defined based on this volume uncertainty $\Delta V$ and the GND $n_\text{gal}$, using the criterion $n_\text{gal} \times \Delta V\leq 1$, rather than a conventional SNR threshold. This criterion is physically motivated since it requires a sufficiently small $\Delta V$ to ensure a unique host identification, which in turn necessitates an exceptionally high-SNR GW signal. Thus, any MBHB meeting this criterion is inherently loud enough to be detected, justifying the replacement of an SNR threshold.



\begin{figure*}[htbp]
    \centering
    \includegraphics[width=1.0\linewidth]{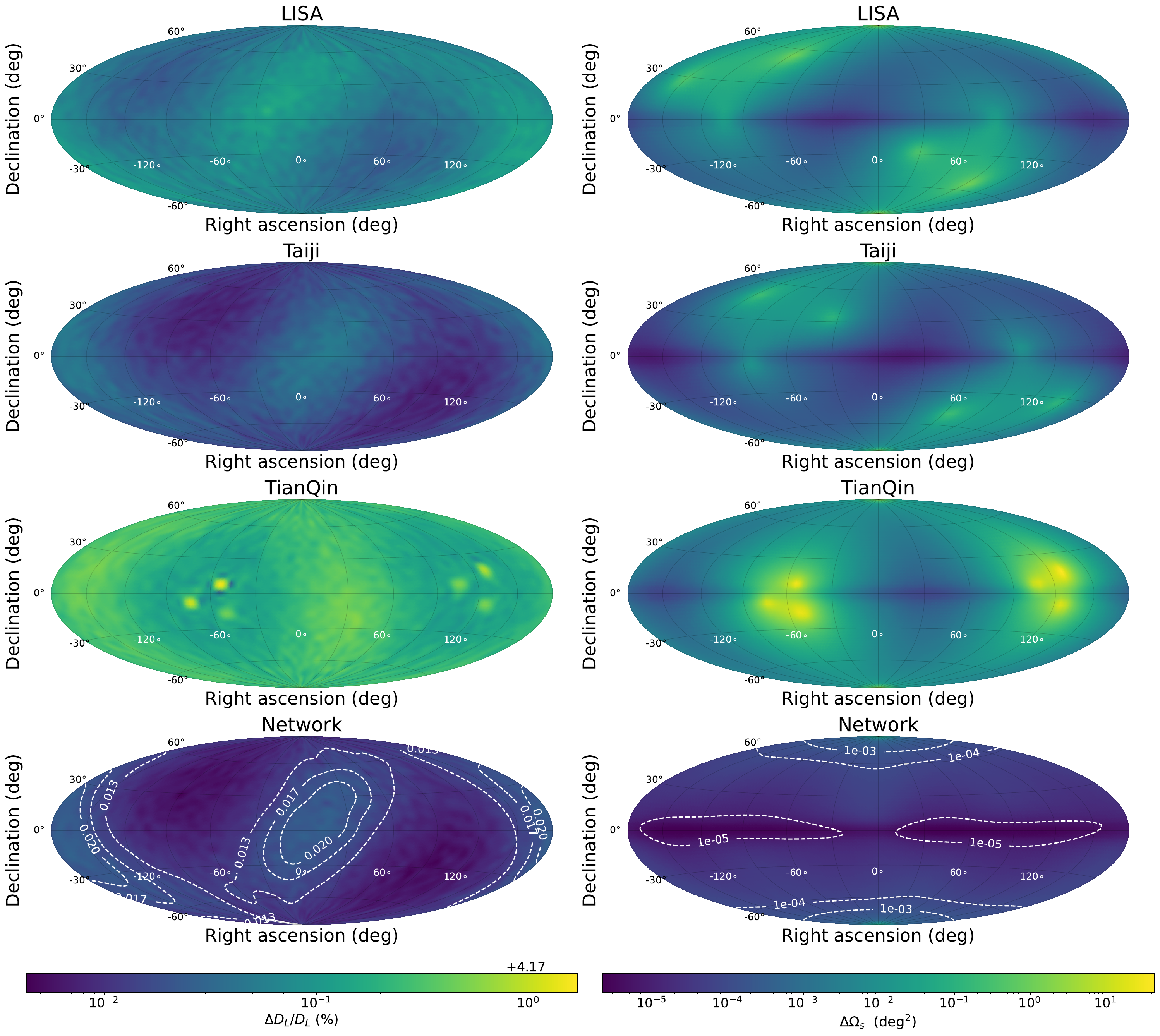}
    \caption{The sky map of fractional distance uncertainty (left column) and sky location uncertainty from equal-mass MBHBs of $10^6M_\odot$ at a distance of $10^4 \text{Mpc}$, comparing single detectors and their network. The top three panels in each column represent the single detectors, and the bottom panel represents their network. The sky maps in each column share the same color bar.
    While the distance uncertainty exhibits only weak sky angular sensitivity, varying by less than an order of magnitude across the sky, the sky location uncertainty is highly sensitive to sky location, varying by factors of $10^2-10^4$. Taiji provides the best constraint on both distances and sky locations, due to its longest arm length. LISA ranks second, while Tianqin has the worst constraint. The network enhances the accuracy of the distance uncertainty by a factor of $2-3$ for a single detector. Additionally, the network highly improves the sky location accuracy by a factor of $\sim 10^2$.}
    \label{fig:dl-dome-angle}
\end{figure*}

\section{Results}\label{sec:result}

\subsection{Joining three GW constellations: Improvements in the sky localization}\label{sec:improvement}

\par In this section, we explore how single GW detectors (LISA, Taiji, TianQin) perform in the accuracy in the sky localization of an MBHB and the improvement achieved by combining them into a network. Detector performance is quantified by the volume uncertainty. A smaller volume uncertainty increases the probability of localizing the host galaxy. According to Eq. \eqref{eq:volume-uncertainty}, the volume uncertainty depends on both the distance uncertainty and the sky location uncertainties. In the following subsections, we compare the network's performance in distance and sky location uncertainties to that of individual detectors using a representative case with specific parameters. In addition, we calculate the maximum distance at which an MBHB can uniquely identify its unique host galaxy using solely GW, across the large parameter space in mass and mass ratio.

\subsubsection{Improvement on distance, sky location and volume uncertainty}\label{sec:distance-location-volume}

\par To assess the localization performance, we examine the angular response of single detectors and their network for MBHBs with equal masses of $10^6M_\odot$ at a distance of $10^4 \text{Mpc}$ without losing generality. Fig. \ref{fig:dl-dome-angle} shows the fractional distance uncertainty (left column) and sky location uncertainty (right column) across the sky. Taiji provides the best constraint on both distances and sky locations, due the its longest arm length. LISA ranks second, while TianQin has the worst constraint. The network enhances the accuracy of the distance uncertainty by a factor of $2-3$ compared to single detectors.  Furthermore, the network highly improves the sky location accuracy by approximately two orders of magnitude $(\sim 10^2)$. The improvement by the network originates from the enhancement of SNR and larger arrival time differences of the GW, which would be discussed in detail in the next subsection. 

\begin{figure*}[htbp]
    \centering
    \includegraphics[width=.9\linewidth]{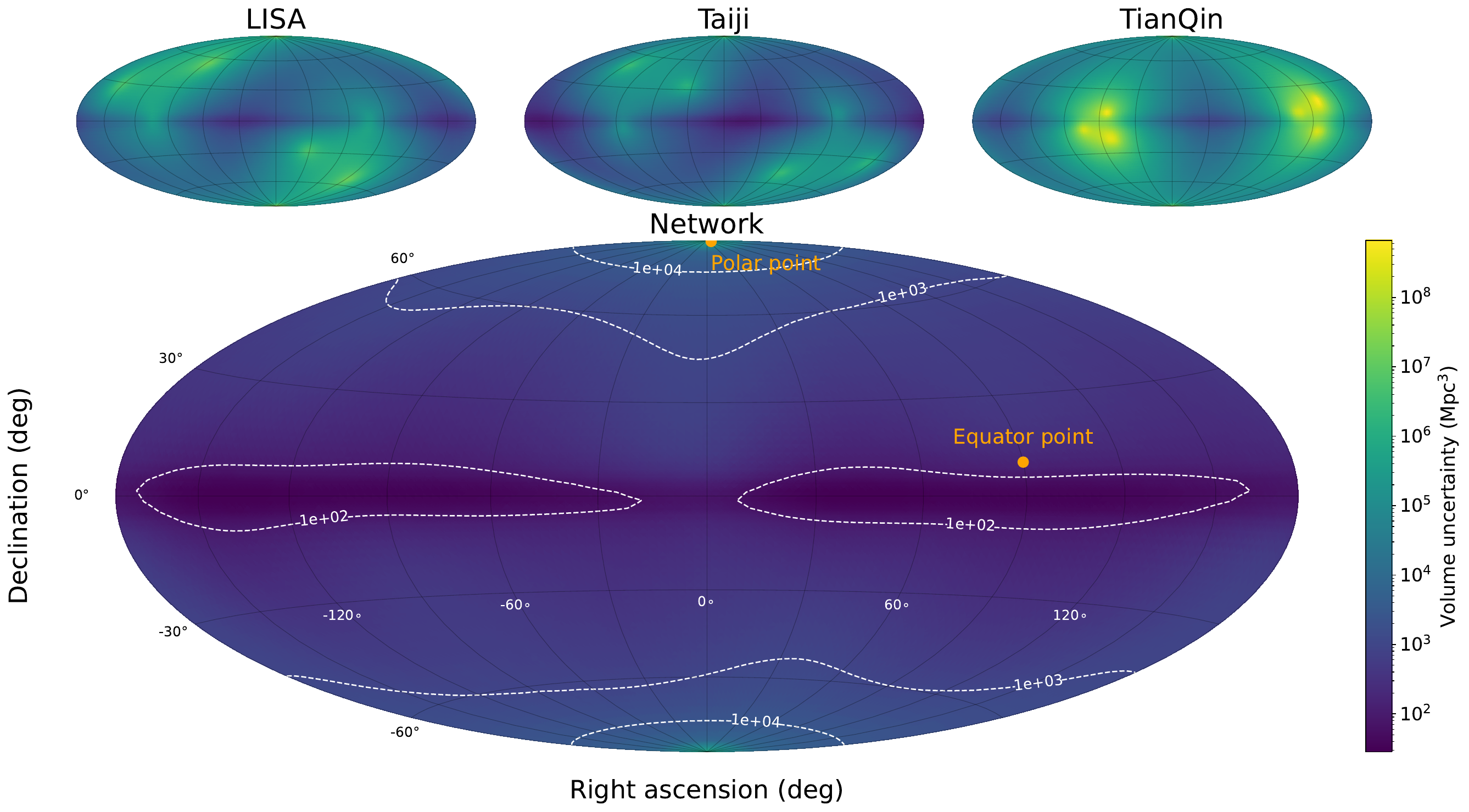}
    \caption{The sky map of the volume uncertainty with an equal mass of $10^6M_\odot$ at a distance of $10^4 \text{Mpc}$, comparing individual detectors and their network. The top panels represent the individual detectors, and the bottom panel represents their network. All the sky maps share the same color bar. The sky location uncertainty reaches its minimum at the celestial equator and its maximum at the celestial pole. We choose the equatorial point $(\theta_s=10^\circ,\phi_s=90^\circ)$ for subsequent calculation, representing the most accurate case. We set the declination $\theta_s=10^\circ$ to prevent the extinction effects from the Galactic plane. Conversely, to represent the most uncertain case, we choose the polar point $(90^\circ,89^\circ)$.}
    \label{fig:vol-angle}
\end{figure*}

\par The distance uncertainty exhibits insensitivity with the sky location, varying less than an order of magnitude across the sky, as it primarily scales with the SNR \citep{takahashiParameterEstimationGalactic2002,kyutokuGravitationalwaveCosmographyLISA2017}. 
In contrast, the sky location uncertainty is highly sensitive to sky position, varying by factors of $10^2-10^4$. Thus, the angular distribution of the volume uncertainty is dominated by the sky location uncertainty. As it is shown in Fig. \ref{fig:vol-angle}, the volume uncertainty matches the angular variation of the sky location uncertainty. As predicted by Eq. \eqref{eq:skylocation}, the sky location uncertainty reaches its minimum at the celestial equator and its maximum at the celestial pole. Correspondingly, the volume uncertainty ranges from $\sim 10^2~\text{Mpc}^3$ at the equator to $\sim 10^8~\text{Mpc}^3$ near the pole. In our following calculation, we choose the equatorial point $(\theta_s=10^\circ,\phi_s=90^\circ)$ for subsequent calculation, representing the most accurate case. We set the declination $\theta_s=10^\circ$ to account for the incompleteness in the galaxy catalogs due to the extinction effects caused by the Milky Way gas. Conversely, to represent the most uncertain case, we choose the polar point $(\theta_s=89^\circ,\phi_s=90^\circ)$.



\subsubsection{Identification horizon under criterion $n_\text{gal}\times \Delta V\leq 1$}\label{sec:horizon}

\par Host galaxies are uniquely identifiable when the number of candidates in the uncertainty volume equals or is less than one. The maximum identification distance, termed the identification horizon, relies on the GND and the localization accuracy of the GW source. \cite{ruanLISATaijiNetworkPrecision2021} predicts a horizon of $z\sim 0.9$ for MBHBs with equal masses of $10^6 M_\odot$ detected by the LISA-Taiji network, assuming a uniform GND of $0.02 \text{ Mpc}^{-3}$. However, the GND evolves across redshift, typically with a lower number density at higher redshift, as demonstrated in Sect. \ref{sec:GND}, potentially providing a further identification horizon. 

\par We simulate the identification horizon of the network across MBHBs with primary masses of $10^5-10^8 M_\odot$ with mass ratios $0.01-1$, covering the sensitivity band of space-based detectors, for both uniform GND and $z$-dependent GND models. Detections from polar and equatorial points are considered, with fixed parameters: 
the dimensionless spins $\chi_1=\chi_2=0.1$, the inclination $\iota=0$, the polarization angle $\psi=1$ rad and the coalescence time $t_c=1$ yr. 


\par The choice of fixed parameters does not affect the generality of identification horizons. Demonstrated in Fig. \ref{fig:para_dependent} in the Appendix, the detection horizon is largely insensitive to variations in $t_c$ and $\psi$  (varying by <1\% and <15\%, respectively), whereas it exhibits stronger dependence on $\iota$, varying by up to $\sim 25\%$. Crucially, any potential bias introduced by these fixed values is irrelevant for the subsequent calculations of detection rates and $H_0$ constraints in Sect. \ref{sec:rate} and \ref{sec:h0}, since the parameters will be fully randomized in that analysis.

\begin{figure*}[htbp]
    \centering
    \includegraphics[width=1\linewidth]{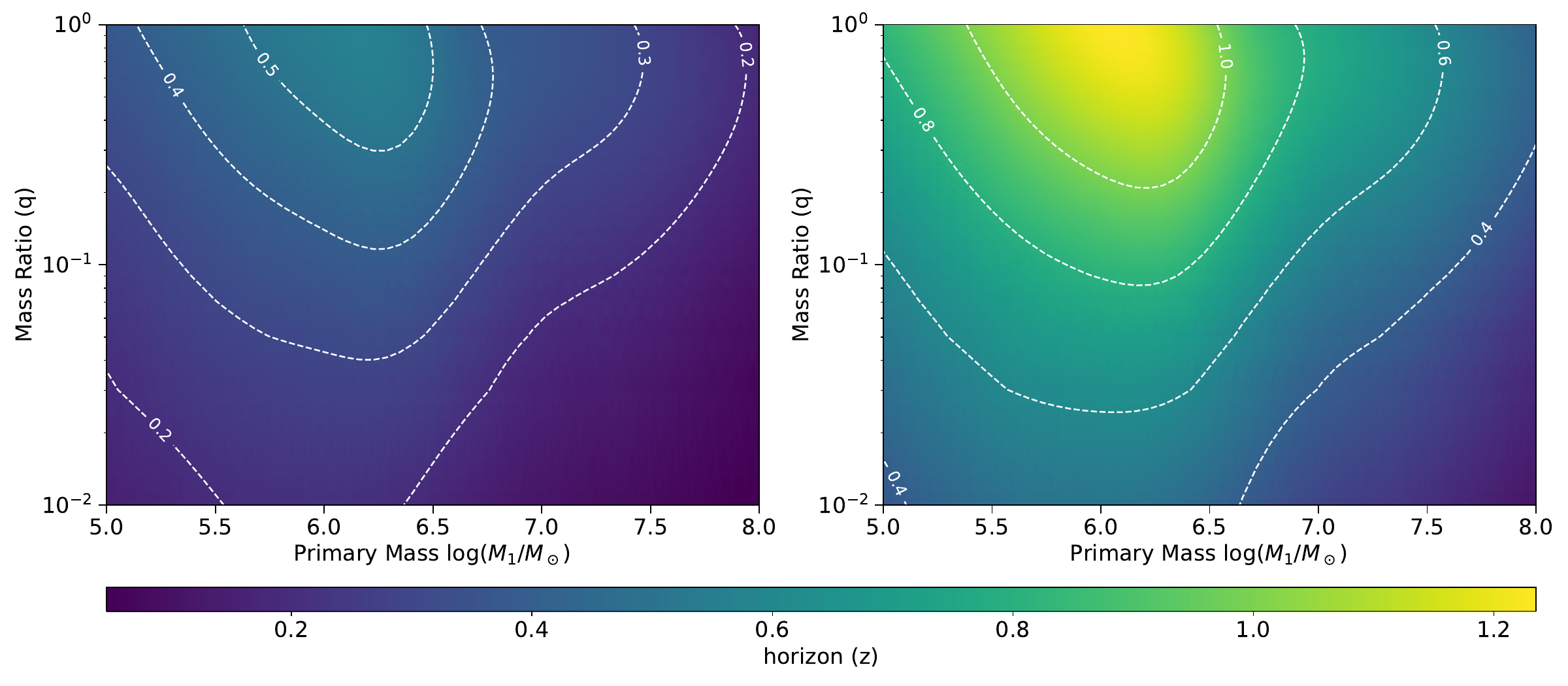}
    \caption{The identification horizon for the \textit{detector network} with the uniform GND ($0.02\text{ Mpc}^{-3}$) at the polar point (left figure) and equatorial point (right figure) for MBHBs with different masses. The maximum identification horizon for both points is reached at the equal-mass system with masses of $\sim 10^6M_\odot$. The identification horizons get closer with a lower or higher mass and a lower mass ratio.}
    \label{fig:horizon-uniformGND}
\end{figure*}

\par
\begin{figure*}[htbp]
    \centering
    \includegraphics[width=1\linewidth]{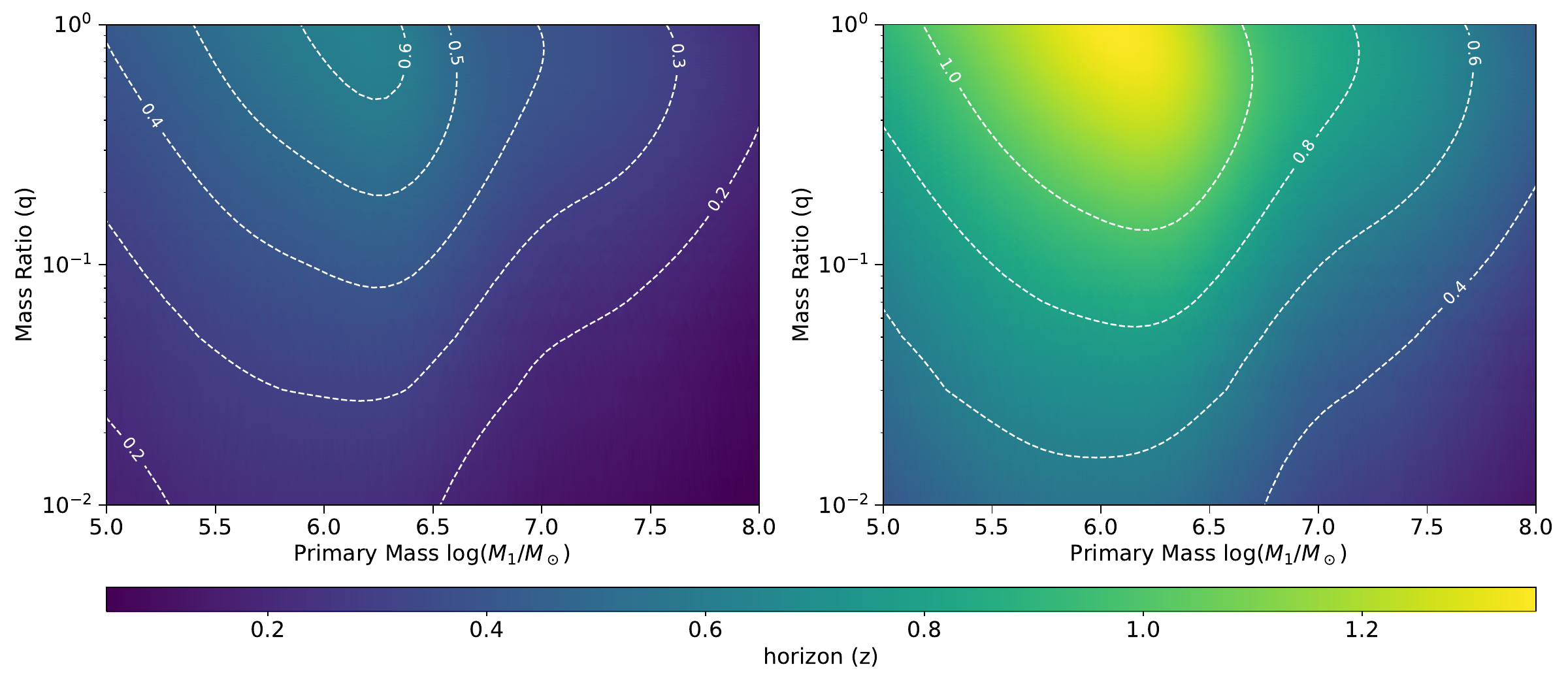}
    \caption{The same as Fig. \ref{fig:horizon-uniformGND} for the \textit{detector network} but with $z$-dependent GND ($n_{8,9}(z),n_{9,14}(z)$). The left figure represents the identification horizon at the polar point, while the right one represents that at the equatorial point. 
    The identification horizon shows similar patterns as Fig. \ref{fig:horizon-uniformGND}, but the horizon is further.}
    \label{fig:horizon-zDGND}
\end{figure*}

 As it's demonstrated in Fig. \ref{fig:horizon-uniformGND} and Fig. \ref{fig:horizon-zDGND}, the network identification horizons detected from the equatorial point exceed those from the polar point by a factor of $\sim 2$ for both GND models. Maximum identification horizons reach in the equal-mass systems with mass around $10^6 M_\odot$, independent of sky location or density models. Horizons decrease for the system with a lower/higher mass and smaller mass ratio due to the decrease in the SNR of the detected MBHB. Specifically, when the mass ratio falls below 0.1, the identification horizon decreases to $z<0.4$ for the uniform GND and to $z<0.8$ for the $z$-dependent GND. 

\par Notably, the identification horizons for single detectors exhibit the same patterns as those from the network (see Fig. \ref{fig:horizon-single-det} in the Appendix for details), though with significantly smaller values. The sky location measurement for a single detector relies on the arrival time differences of GW between constituent satellites, whereas for a detector network, it is based on the time differences between different detectors. Consequently, employing a detector network reduces the sky location uncertainty by more than two orders of magnitude, as demonstrated in Fig. \ref{fig:dl-dome-angle}. In contrast, the distance uncertainty is limited when using the network. Theoretically, the fractional error on distance is inversely proportional to the SNR, given by $(\Delta {d_L}/d_L)\propto \rho^{-1}$ \citep{takahashiParameterEstimationGalactic2002,kyutokuGravitationalwaveCosmographyLISA2017}. The detector network could lead to enhanced SNRs by a factor $\sim 2$, calculated as $\rho=\sqrt{ \rho_{\text{LISA}}^2+\rho_{\text{Taiji}}^2+\rho_{\text{TianQin}}^2}$. Additionally, the uncertainty due to the weak lensing effect becomes dominated at a large distance of $\sim 5\times 10^3 \text{ Mpc}$, further diminishing the network’s improvement on distance measurements. Therefore, the enhancement of the identification horizon by the detector network primarily lies in the reduction in sky location uncertainty.
\par For a single detector, even in the most optimistic scenario with $z$-dependent GND detecting $\sim 10^6 M_\odot$ MBHBs from the equatorial point, the identification horizon remains less than $z\sim 0.6$. With the uniform GND, the horizon is even restricted, at $z<0.4$. In contrast, the detector network increases the identification horizon by a factor of approximately $2$ compared to single detectors.

\subsection{The statistics of MBHBs as standard sirens: Detection rates }\label{sec:rate}

\par In this section, we utilize simulated MBHB catalogs generated from population models to investigate their potential as "bright" siren but without EMc and explore the possibility of using these detections to constrain the $H_0$. 

\par Five simulated catalogs of MBHBs are generated for five population models under a 4-year observation scenario. We exclude MBHBs at redshifts $z > 3$, as \cite{mangiagliObservingInspiralCoalescing2020} demonstrated that beyond this redshift, the sky localization becomes sufficiently large to hinder effective sky-localization searches. The upper limit also arises because the identification horizon, shown in Fig. \ref{fig:horizon-uniformGND} and Fig. \ref{fig:horizon-zDGND}, lies at $z \sim 1$. In addition, we apply a mass cutoff to the MBHB populations, requiring at least one black hole in the binary to have a mass $>10^5 M_\odot$ to exclude low-mass dwarf galaxies, whose high GND significantly prevents effective EMc-searching. The parameters of the MBHBs are drawn from each population or randomized: The binary mass $(m_1,m_2)$, luminosity distance $d_L$ are extracted from the population, while other parameters $(\iota,\psi, t_c,\theta_s,\phi_s)$ are uniformly randomized, with $(\iota,\psi,\theta_s)\in [-\pi/2,\pi/2]$, $\phi_s\in [0,2\pi]$ and $t\in [0,4]\text{ yr}$. The spin of the MBHB is fixed, i.e. $\chi_1=\chi_2=0.1$, for simplicity. The sources near the galactic plane ($|\theta_s| <10^\circ$) are excluded due to the extinction effects of the Galactic plane. Each catalog undergoes 1000 realizations through statistical resampling.  We evaluate the detectability of each MBHB individually. Only those within identification horizons satisfying ($n_\text{gal} \times \Delta V \leq 1$) permit unique host galaxy identification and thus qualify as bright sirens. We emphasize that detection is determined by this horizon criterion, not by a typical SNR threshold.



\begin{table*}[htbp]
\caption{The detection rate and averaged constraint on $H_0$ for 1000 realizations} \label{tab:h0}
\centering 
\hspace{-3.cm}
\resizebox{1.1\textwidth}{!}{%
\begin{tabular}{lccccccccc}
\hline
\multirow{2}{*}{Model} & \multirow{2}{*}{\parbox{2cm}{~\\ \centering Event  rate \\ (yr$^{-1}$)}} & \multicolumn{2}{c}{\parbox{2.5cm}{\centering Realizations \\ with detections}} & \multicolumn{1}{c}{} & \multicolumn{2}{c}{\parbox{3cm}{\centering Detection rate \\ (yr$^{-1}$)}} & \multicolumn{1}{c}{} & \multicolumn{2}{c}{\parbox{4cm}{~\\\centering Averaged $H_0$ constraint\\ $\Delta H_0/H_0 (\%)$}} \\ \cline{3-4} \cline{6-7} \cline{9-10} 
 &  & \parbox{1.5cm}{\centering ~\\uniform\\GND\\ \tiny~} & \parbox{2cm}{\centering ~\\$z$-dependent\\GND\\ \tiny~} & \multicolumn{1}{c}{} & \parbox{1.5cm}{\centering ~\\uniform\\GND\\ \tiny~} & \parbox{2cm}{\centering ~\\$z$-dependent\\GND\\ \tiny~} &  & \parbox{1.5cm}{\centering ~\\uniform\\GND\\ \tiny~} & \parbox{2cm}{\centering ~\\$z$-dependent\\GND\\ \tiny~} \\  \hline 
LightSeed-Nodelay & 166 & 190 & 204 &&$0.057\pm 0.127$ & $0.061 \pm 0.130$   & & 1.02$\pm$0.90 & 1.14$\pm$1.00   \\
HeavySeed-Delay & 8 & 286 & 348 &  &$0.077\pm 0.127$  &  $0.109 \pm 0.164$ & & 1.74$\pm$0.60 & 1.88$\pm$0.80\\
HeavySeed-Nodelay & 120 & 138 & 197 && $0.035\pm 0.090$  &  $0.053 \pm 0.110$ &  & 1.75$\pm$0.67 & 2.04$\pm$1.28 \\
IV24 & 6 & 966 & 971 &  &$0.866\pm 0.464$  &  $0.940 \pm 0.494$ &  & 0.76$\pm$1.62 & 0.70$\pm$1.13 \\
QY25 & 6 & 927 & 967 & & $0.684\pm 0.412$  &  $0.879 \pm 0.480$ &   & 0.80$\pm$0.73 & 0.76$\pm$0.68 \\
 \hline
\end{tabular}}
\end{table*}

\par The results are summarized in Table \ref{tab:h0}. For the LightSeed-Nodelay model, only 19\% of realizations (190 of 1000)  of all yield detections with the uniform GND, whereas the $z$-dependent GND slightly improves detectability, achieving the percentage to 20\%. The HeavySeed-Delay and HeavySeed-Nodelay models exhibit similar behavior, with 29\% and 14\% realizations yielding detections in uniform GND, respectively. These percentages increase to 35\% and 20\% in the $z$-dependent GND. Conversely, the IV24 and QY25 models ensure in $>90\%$ of realizations for both GND models. The detection percentage for the IV24 increases slightly from 95\% (uniform) to 96\% ($z$-dependent), while QY25 increases from 94\% to 97\%.

\par We compute averaged detection rates and their standard deviations across 1000 realizations for all models. For LightSeed-Nodelay, HeavySeed-Delay, and HeavySeed-Nodelay models, uniform GND yields detection rates below $0.1 \text{ yr}^{-1}$, indicating pessimistic prospects for detection within a 4-year observation. The $z$-dependent GND marginally improves these rates, but the detectability remains low. Conversely, the IV24 and QY25 models exhibit significantly higher detection rates of 0.866 $\pm$ 0.464  yr$^{-1}$ and 0.684 $\pm$ 0.412 yr$^{-1}$ under the uniform GND. These rates increase to  $\sim 0.9$ yr$^{-1}$ with $z$-dependent GND, substantially enhancing detection prospects.

\par There are differences in detectability between different population models and GND models. As demonstrated in Fig. \ref{fig:population}, the LightSeed-Nodelay model favors MBHB population at high redshifts ($z>4$), and low-mass MBHBs (binary mass lower than $10^5M_\odot$). However, given the identification horizon of $z\sim 1.2$ in the most accurate case (detected at the equatorial point), few MBHBs in this model could be detected within the identification horizon. Furthermore, we apply a mass cutoff to the MBHB populations, requiring at least one black hole in the binary to have a mass $>10^5 M_\odot$ to exclude low-mass dwarf galaxies. This cutoff eliminates a substantial fraction of the population, further reducing detectable events. The HeavySeed-Nodelay model faces the same challenge in redshift distributions. This model shows a dominance of MBHBs with masses in the range of $10^6-10^7M_\odot$, well above the mass cutoff. However, the mass ratio of the population is largely uniformly distributed over the range of 0.1 to 1, leading to a closer horizon. Consequently, both the LightSeed-Nodelay and HeavySeed-Nodelay models exhibit high event rates but low detection rates. Similarly, the HeavySeed-Delay model features an MBHB population spanning a broad redshift range $(z=2-6)$, leading to low detection rates despite the event rate peaks at binary mass around $10^6M_\odot$. Conversely, the population in the IV24 and QY25 models concentrates at low redshifts $(z<2)$, indicating a high detection possibility. The mass cutoff has minimal impact on these two models due to favorable redshift distributions. Thus, an appropriate MBHB population model is essential for accurately forecasting future detections and their cosmological implications. The future detections also benefit in building appropriate population models or improving current models.

\par Moveover, the GND model also affects the detection rates. Compared to the uniform GND of $0.02 \text{ Mpc}^{-3}$, the $z$-dependent GND slightly improves the detection rates as shown in the result (Sect. \ref{sec:result}). Demonstrated in Fig. \ref{fig:GND}, the components $n_{8,9}(z)$ (for MBHBs with binary mass in range of $10^5-10^6M_\odot$; defined in Sect. \ref{sec:GND}) and $n_{9,14}(z)$ (for MBHBs with binary mass above $10^6M_\odot$) decreases by a factor of $\sim 0.3-0.4$ at $z\sim 1.2$ compared to $0.02 \text{ Mpc}^{-3}$. This extends the detection horizon, improving source localization and increasing detectable events. Notably, the two components in the $z$-dependent GND in this work are roughly chosen for general situations, simply based on the relations between the stellar mass of the galaxy and the mass of the central black hole. A refined $z$-dependent GND may improve the predicted detection rates. 

\subsection{The statistics of MBHBs as standard sirens: Cosmological constraints}\label{sec:h0}

\par
\begin{figure*}[htbp]
    \centering
    \subfigure[Constraint on $H_0$ with uniform GND]{\includegraphics[width=0.45\linewidth]{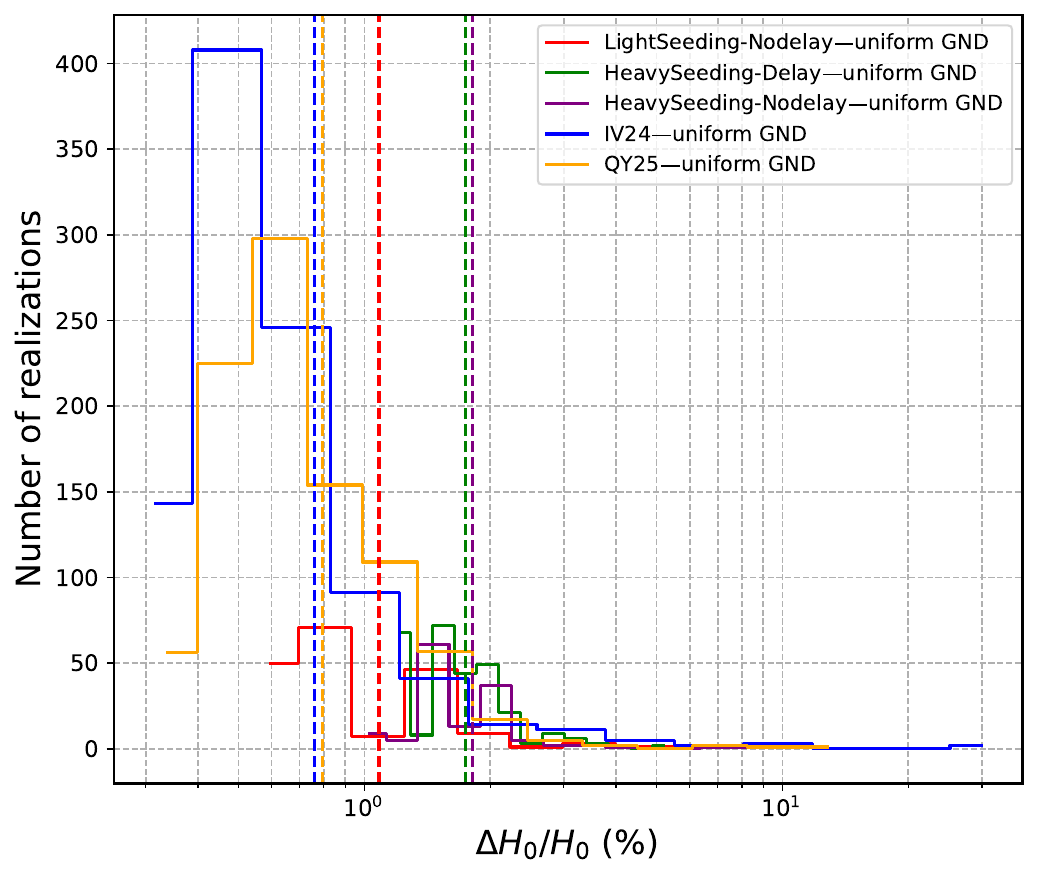}}\label{fig:h0-constantGND}
    \subfigure[Constraint on $H_0$ with $z$-dependent GND]{\includegraphics[width=0.45\linewidth]{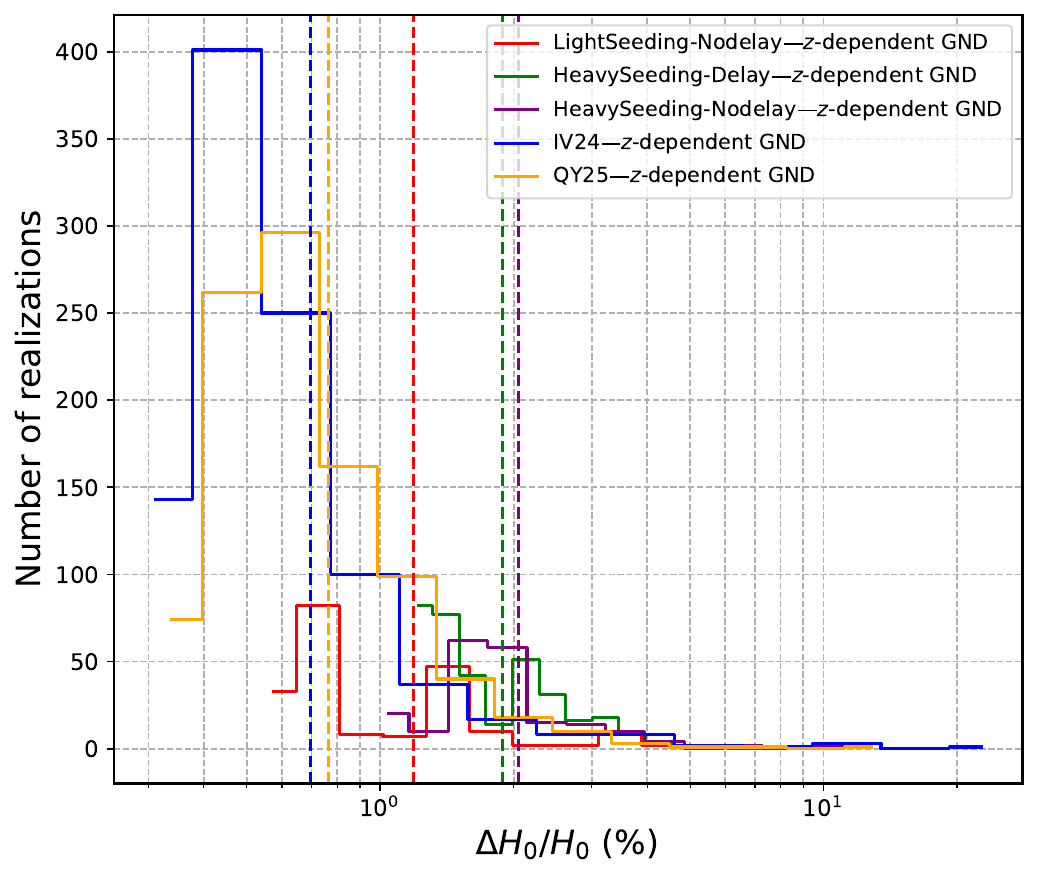}}\label{fig:h0-zdependentGND}
    \caption{The distribution of fractional uncertainty on $H_0$ across realizations. The averaged uncertainties for each population model are plotted in colored dashed lines. Fig. (a) represents the constraints with the uniform GND. The fractional uncertainty for the LightSeed-Nodelay, HeavySeed-Delay and HeavySeed-Nodelay models spans a range of approximately $0.6-10\%$ with uniform GND, whereas the QY25 model produces tighter constraints, spanning 0.3–6\%. In contrast, the IV24 model exhibits significantly greater variation under both GND assumptions, with most realizations constraining $H_0$ to better than 1\%, though a small number exceed 10\%.}  
    \label{fig:h0}
\end{figure*}

 We assume that all the MBHBs within the identification horizon can be unambiguously associated with their host galaxies. These sources are assumed as detected bright sirens to constrain the $H_0$. In a single realization, all detections are assumed to be independent and jointly contribute to the constraint on $H_0$. The joint constraints are averaged across realizations \textit{contented} detections. In Table \ref{tab:h0}, we present the averaged fractional uncertainty on $H_0$ of realizations, namely, $\Delta H_0/H_0$ (where $\Delta H_0$ denotes the uncertainty with $1\sigma$ credible interval). The uncertainty distribution across realizations for both GNDs is illustrated in Fig. \ref{fig:h0}. For LightSeed-Nodelay, HeavySeed-Delay, and HeavySeed-Nodelay models, the fractional uncertainties span a range of approximately $0.6-10\%$ with uniform GND, with averaged fractional uncertainties of 1.02$\pm$0.90\%, 1.74$\pm$0.60\% and 1.75$\pm$0.67\%, respectively. But the uncertainties gently degrade to 1.14$\pm$1.00\%, 1.88$\pm$0.80\% and 2.04$\pm$1.28 under the $z$-dependent GND, as demonstrated in Table \ref{tab:h0}. This degradation stems from the extended identification horizon under $z$-dependent GND: while the extended horizon enables more distant MBHB detections, these sources provide poorer $H_0$ constraints due to the decrease of the SNR of the detected MBHB. In contrast, the uncertainties for the QY25 model span a lower range of $0.3 - 6\%$. The constraints for the IV24 model under both GNDs vary more dramatically, with most realizations yielding a constraint below $1\%$ and a small number of realizations above 10\%. The IV24 and QY25 models achieve the average $\Delta H_0/H_0 < 1\%$ for both GNDs. Notably, both GNDs yield similar constraints for the two population models. Generally, the above constraints are better than those from other probes at present \citep{yuHubbleParameterBaryon2018,wu8CentDetermination2022}.

\subsection{Network enhancement for detection rates}

\par Illustrated in Sect. \ref{sec:improvement}, the network significantly enhances the identification horizon. The limited identification horizon of single detectors results in few or no detections, as most population models predict MBHBs at high redshift ($z>1$). As shown in Table \ref{tab:single-detector-number} in the Appendix, only dozens or even no realizations out of 1000 are detected MBHBs within the identification horizons for LightSeed-Nodelay, HeavySeed-Delay and HeavySeed-Nodelay models with single detectors (LISA, Taiji, or TianQin). The $z$-dependent GND improves the detectability, but the fraction of realizations with detections remains low ($< 4\%$). In particular, the LightSeed-Nodelay yields no detections with any single detector. To make matters worse, most realizations with detections contain only few events, yielding very low detection rate ($\lesssim 10^{-3}\text{ yr}^{-1}$). Even for the most refined models, such as IV24 and QY25, only $>50\%$ realizations yield detections with both GNDs. The rates remain below $0.2$~yr$^{-1}$, implying an uncertain detection prospect over a 4-year observation period with single detectors. Thus, the network is essential for identifying host galaxies of MBHBs without EMc.

\subsection{Comparison with the next generation GW detectors}

\par Identifying the host galaxy with solely the GW signal gets rid of the dilemma of searching for EMc. These GW sources are promising standard sirens to constrain $H_0$. This method relies on ultra-high-precision localization of the GW sources. In this work, we discuss the detectability by this method via the space-based GW detectors and their network. However, other GW detectors may also exhibit the capability of host galaxy identification.

\par The LIGO-Virgo-KAGRA (LVK) network has shown progressive improvement in sky localization from O1 to O4 runs, with sky location uncertainty reducing from $\sim 1000$ deg$^2$ to dozens of deg$^2$ \citep{ligoscientificcollaborationandvirgocollaborationGWTC1GravitationalWaveTransient2019,ligoscientificcollaborationandvirgocollaborationGWTC2CompactBinary2021,ligoscientificcollaborationGWTC3CompactBinary2023,collaborationGWTC40UpdatingGravitationalWave2025}. Projections for the forthcoming O5 run indicate typical sky location uncertainty remaining at the level of tens of deg$^2$ for best localized detections \citep{kiendrebeogoUpdatedObservingScenarios2023}.
Given a typical distance uncertainty of $10\%$ for the LVK network and a uniform GND of $0.1$ Mpc$^{-3}$ at low redshift, the identification horizon to identify a unique host galaxy for the LVK network should be approximately 20 Mpc. Observations constrain the event rate of stellar mass compact binary mergers in the local universe to around 1000 Gpc$^{-3}$ yr$^{-1}$ \citep{abbottRATEBINARYBLACK2016, ligoscientificcollaborationandvirgocollaborationGWTC1GravitationalWaveTransient2019}. With this event rate, the LVK network may detect sources within the identification horizon in a decade. However, the uncertainty from peculiar velocity of the host galaxy dominates at such a low redshift. Thus, the constraint on $H_0$ by the LVK network with this method would be inaccurate without sufficient sources.

\par Next-generation ground-based detectors Cosmic Explorer \citep[CE;][]{reitzeCosmicExplorerUS2019} and Einstein Telescope \citep[ET;][]{punturoEinsteinTelescopeThirdgeneration2010} are expected to achieve sub-degree localization, with sky location uncertainty below 10 deg$^2$ at low redshift $\sim 0.2$ \citep{zhaoLocalizationAccuracyCompact2018,nitzPremergerLocalizationCompactbinary2021}. The distance uncertainty may achieve $5\%$. Thus, the identification horizon for CE and ET is $\sim 40-90$ Mpc, leading to potential detections within the horizon in a few years. 

\par Space-based detectors like Big Bang Observer \citep[BBO;][]{harryLaserInterferometryBig2006} and Deci-hertz Interferometer Gravitational Wave Observatory \citep[DECIGO;][]{kawamuraJapaneseSpaceGravitational2011} promise exceptional precision due to longer arm length. These detectors are expected to localize stellar mass compact binary mergers to $\sim 1$ deg$^2$ with $\sim 1\%$ distance uncertainty \citep{cutlerUltrahighPrecisionCosmology2009}. The localization for massive binary mergers may be even more accurate due to stronger signals. \cite{cutlerUltrahighPrecisionCosmology2009} discussed that BBO could be able to uniquely identify the host galaxies of stellar mass compact binary mergers within $z\sim 5$. Decigo may achieve similar detections. 

\par These advancements will significantly enhance multi-messenger astronomy and cosmological studies, cooperating with the LISA-Taiji-TianQin network.

\section{conclusion}\label{sec:conclusion}

This work studies the potential of localizing the host galaxy of an MBHB without EMc, using the ultra-high-precision localization of the space-based GW detector network, LISA-Taiji-Tianqin. Such MBHBs can serve as bright sirens to provide constraints on $H_0$. The accuracy of host galaxy localization depends on both the MBHB populations and the GND models. We consider two GND models: a uniform model with constant density of $0.02 \text{Mpc}^{-3}$, and a $z$-dependent model composed of two evolving components. In addition, we simulate catalogs for five MBHB population models, i.e., the LightSeed-Nodelay, HeavySeed-Nodelay, HeavySeed-Delay, IV24 and QY25, each with 1000 realizations. MBHB detection is determined by the GND $n_{\text{gal}}$ and the measured volume uncertainty of the MBHB $\Delta V$, using the criterion $n_\text{gal}\times \Delta V\leq 1$. The MBHB populations are filtered by a mass cutoff requiring that at least one in the binary has a mass greater than $10^5M_\odot$ to exclude the low-mass dwarf galaxies. Moreover, we exclude the population co-planar with the Galactic plane ($\theta_s<10^\circ$) to prevent the extinction effects. We calculate the detection rates of MBHBs within the horizon for all population models and GND models. We estimate the constraint on $H_0$ with the detected MBHBs.
\par The main results are summarized as follows:
\begin{enumerate}

    \item the network enhances the accuracy of the distance uncertainty of an MBHB by a factor of $2-3$ compared to single detectors, while enhancing the sky location accuracy by a factor of $10^2$. The angular dependence of the localization volume uncertainty is dominated by sky position uncertainty, which is minimized at the celestial equator and maximized at the celestial poles.

    \item The identification horizon reaches a maximum of $z\sim 1.2$ for equal-mass ($10^6 M_\odot$) MBHBs in the most accurate localization case (i.e., detected at the equatorial point) for both GND models. The $z$-dependent GND gently increases the horizon. The horizon gets closer with a lower or higher mass and a lower mass ratio due to the decrease in the SNR of the detected MBHB. 
    
    \item Averaged over 1000 realizations, the LightSeed-Nodelay, HeavySeed-Nodelay, HeavySeed-Delay models exhibit pessimistic detection rates $<0.1 \text{ yr}^{-1}$. In contrast, the IV24 and QY25 models show significantly higher detection rates of 0.866$\pm$0.464 and 0.684$\pm$0.412, respectively. Moreover, the $z$-dependent GND further enhances the rate to 0.940$\pm$0.494 and 0.879$\pm$0.480. These higher rates suggest a strong prospect for detections within a 4-year observation period. 
    \item For realizations with detections, the LightSeed-Nodelay, HeavySeed-Delay and HeavySeed-Nodelay models yield the fractional uncertainties on $H_0$ ($\Delta H_0 /H_0$) in a range of approximately $0.6-10\%$ with uniform GND, with averaged fractional uncertainties of 1.02$\pm$0.90\%, 1.74$\pm$0.60\% and 1.75$\pm$0.67\%, respectively. The uncertainties gently degrade to 1.14$\pm$1.00\%, 1.88$\pm$0.80\% and 2.04$\pm$1.28 under the $z$-dependent GND. The IV24 and QY25 models provide more accurate constraint with averaged $\Delta H_0 /H_0<1\%$ for both GND models: IV24 yields 0.76$\pm$1.62\% (uniform) and 0.70$\pm$1.13\% ($z$-dependent), while QY25 yields 0.80$\pm$0.73\% (uniform) and 0.76$\pm$0.68\% ($z$-dependent). These results indicate a promising pathway for high-precision $H_0$ measurement. 
    \item Detection rates for individual detectors (LISA, Taiji, or Tianqin) are  very low ($\lesssim 10^{-3} \text{ yr}^{-1}$) for the LightSeed-Nodelay, HeavySeed-Delay and HeavySeed-Nodelay models. For the IV24 and QY25 models, rates remain below $0.2\text{ yr}^{-1}$, implying an uncertain detection prospect over a 4-year observation period with single detectors. Hence, the detection network is essential to localize a host galaxy without EMc.
\end{enumerate}







\section*{Acknowledgement}
We thank Muchun Chen and Yiming Hu for their useful discussion on the science of GW. We also thank Jianping Hu and Xuandong Jia for their discussion. We thank Silvia Bonoli for her contribution to the IV24 model. This work was supported by the National Natural Science Foundation of China (grant Nos. 12494575 and 12273009). Y.J.Z. is supported by the Postgraduate Research \& Practice Innovation Program of Jiangsu Province KYCX25\_0198. D.I.V acknowledges the financial support provided under the European Union’s H2020 ERC Consolidator Grant ``Binary Massive Black Hole Astrophysics'' (B Massive, Grant Agreement: 818691) and the European Union Advanced Grant ``PINGU'' (Grant Agreement: 101142079). X.G. is supported by the fellowship of China National Postdoctoral Program for Innovative Talents (Grant No. BX20230104). 

\appendix


\section{Detections with single detector}\label{sec:detection-single-detector}

\begin{figure}[htbp]
    \centering
    \includegraphics[width=0.9\textwidth]{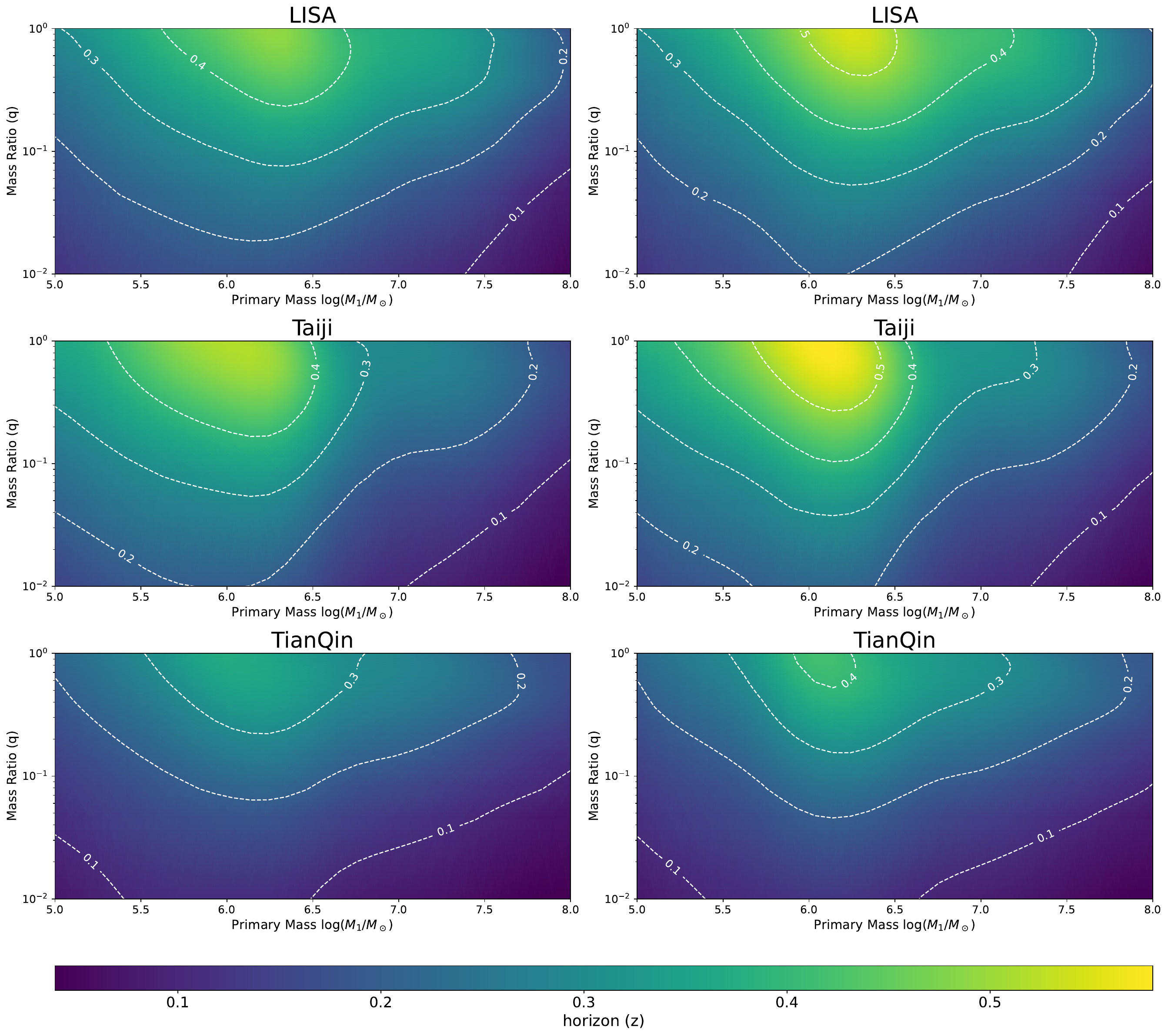}
    \caption{The identification horizons from single detectors at equatorial points with the uniform GND (left) and $z$-dependent GND (right).  All figures share the same color bar. The identification horizons show the same pattern as Fig. \ref{fig:horizon-uniformGND} and Fig. \ref{fig:horizon-zDGND}. Taiji has the farthest identification horizon, and LISA ranks second, followed by Tianqin, which consists of the result of volume uncertainty.}
    \label{fig:horizon-single-det}
\end{figure}
\par We simulate the identification horizon for single detectors, LISA, Taiji, or TianQin at the equatorial point as a reference for the upper detection limit (MBHBs detected at equatorial point). Demonstrated in Fig. \ref{fig:horizon-single-det}, the identification horizon is less than $z\sim 0.4$ with uniform GND, while the $z$-dependent GND expands the horizon for each detector by a factor of $\sim 1.5$. The identification horizons exhibit the same pattern as those with the detector network (shown in Fig. \ref{fig:horizon-uniformGND} and Fig. \ref{fig:horizon-zDGND}).

\begin{table}[htbp]
\centering
\caption{The averaged detection rates of single detectors for 1000 realizations}\label{tab:single-detector-number}

\hspace{-3.cm}
\resizebox{1.1\textwidth}{!}{%
\begin{tabular}{@{}lcccccccclccclccc@{}}
\toprule
\multirow{2}{*}{Model} & \multirow{2}{*}{} & \multicolumn{3}{c}{\parbox{4cm}{\centering Realizations with detections\\uniform GND}} &  & \multicolumn{3}{c}{\parbox{4cm}{\centering Realizations with detections\\$z$-dependent GND}} &  & \multicolumn{3}{c}{\parbox{4cm}{\centering Detection rate (yr$^{-1}$)\\uniform GND}} &  & \multicolumn{3}{c}{\parbox{4cm}{\centering Detection rate (yr$^{-1}$)\\$z$-dependent GND}} \\ \cmidrule(lr){3-5} \cmidrule(lr){7-9} \cmidrule(lr){11-13} \cmidrule(l){15-17}   
 &  & LISA & Taiji & TianQin &  & LISA & Taiji & TianQin &  & LISA & Taiji & TianQin &  & LISA & Taiji & TianQin \\ \midrule
LightSeed-Nodelay &  & 8 & 24 & 1 &  & 15 & 37 & 1 &  & $0.002\pm 0.022$ & $0.006\pm 0.038$ & $<10^{-3}$ &  & $0.004 \pm 0.030$ & $0.009 \pm 0.047$ & $<10^{-3}$ \\
HeavySeed-Delay &  & 1 & 2 & 0 &  & 1 & 5 & 1 &  & $<10^{-3}$ & $0.001\pm 0.011$ & - &  & $<10^{-3}$ & $0.001 \pm 0.018$ & $<10^{-3}$ \\
LightSeed-Nodelay &  & 0 & 0 & 0 &  & 0 & 0 & 0 &  & - & - & - &  & - & - & - \\
IV20 &  & 359 & 452 & 30 &  & 408 & 492 & 48 &  & $0.258\pm 0.261$ & $0.277\pm 0.183$ & $0.081\pm 0.143$ &  & $0.283 \pm 0.274$ & $0.325 \pm 0.287$ & $0.090 \pm 0.150$ \\
QY25 &  & 103 & 152 & 8 &  & 147 & 198 & 9 &  & $0.028\pm 0.084$ & $0.039\pm 0.100$ & $0.004\pm 0.034$ &  & $0.036 \pm 0.095$ & $0.057 \pm 0.119$ &  $0.006 \pm 0.037$ \\ \hline
\end{tabular}%
}
\end{table}

\par We also quantify detections for single detectors using the same catalogs described in Sect. \ref{sec:rate}, as shown in Table \ref{tab:single-detector-number}. For LightSeed-Nodelay, HeavySeed-Delay and HeavySeed-Nodelay models, only several or even no realizations out of 1000 (a fraction $\lesssim 1\%$) detect MBHBs within the identification horizons. Among single detectors, Taiji achieves the best detectability with $z$-dependent GND for the three population models, yet fewer than 4\% of its realizations contain detections. LISA and TianQin perform worse, with the ratio of $\leq 1\%$ or even $\leq 0.1\%$. Moreover, most realizations with detections contain only a few events, yielding very low detection rates ($\lesssim 10^{-3}$~yr$^{-1}$). Even for most refined models such as IV24 and QY25, detections occur in $\lesssim 50\%$ of realizations under both GNDs. The detection rates are below $0.1$ yr$^{-1}$ for the QY25 model, leading to no detections over several years of observation with single detectors. Though detection rates reach $\lesssim 0.3$ yr$^{-1}$ for the IV24 model, detection remains uncertain. In all cases, single-detector performance is substantially poorer than that of the network. 

\section{parameter dependence of the identification horizon}\label{sec:para-dependence}

\begin{figure}[htbp]
    \centering
    \includegraphics[width=0.8\linewidth]{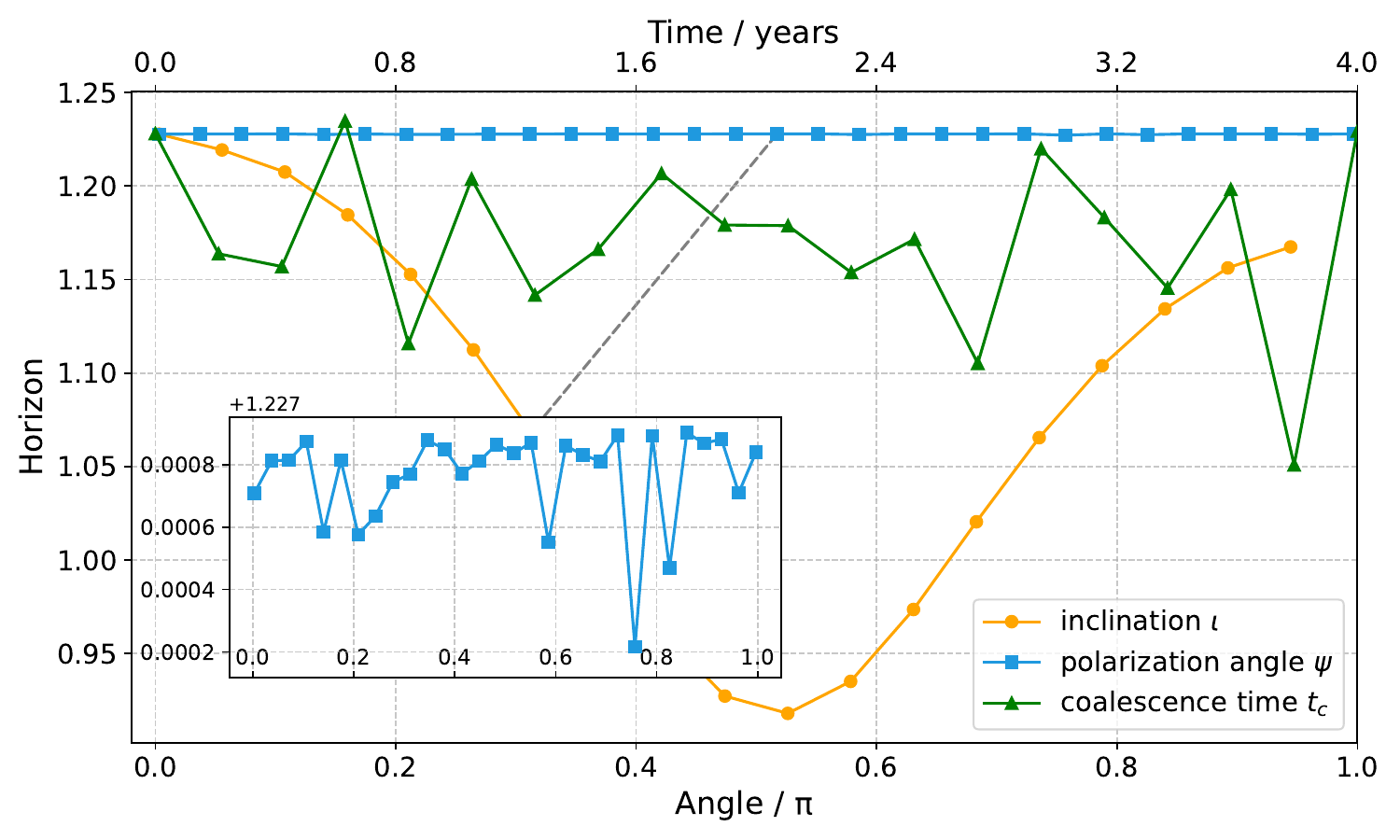}
    \caption{The dependence of the identification horizons under uniform GND of 0.02 Mpc$^{-3}$ on inclination angle $\iota$ (orange line), the polarization angle $\psi$ (blue line), and the coalescence time $t_c$ (green line). The inset provides a magnified view of the blue curve. The horizon is most sensitive to $\iota$, reaching a maximum of $z \sim 1.2$ at $\iota = 0$ and a minimum of $z \sim 0.95$ at $\iota = \pi/2$ (changed by 25\%). In comparison, the horizon shows significantly weaker dependence on $t_c$, varying by less than 15\%. It is even less sensitive to $\psi$, with variations confined to less than 1\%.}
    \label{fig:para_dependent}
\end{figure}

The main result of this paper demonstrates the identification horizon of MBHBs in different mass configurations with a fixed parameter set where $\iota=0$, $\psi=1 \text{ rad}$ and $t_c=1 \text{ yr}$. These three parameters affect the identification horizon. The Fig. \ref{fig:para_dependent} demonstrates identification horizons under uniform GND of 0.02 Mpc$^{-3}$ changed along these three parameters. The horizon is most sensitive to $\iota$, reaching a maximum of $z \sim 1.2$ at $\iota = 0$ and a minimum of $z \sim 0.95$ at $\iota = \pi/2$ (a variation of $\sim 25\%$). In comparison, the horizon shows significantly weaker dependence on $t_c$, varying by less than 15\%. It is even less sensitive to $\psi$, with variations confined to less than 1\%. We emphasize that the fixed parameters would \textit{not} affect the results of the detection rates and $H_0$ constraints (demonstrated in Sect. \ref{sec:h0}), since parameters of the MBHBs in the catalogs are completely randomized.

\bibliography{Cosmology}{}
\bibliographystyle{aasjournal}                                                                                                           
\end{document}